\documentclass[aps,prb,twocolumn,showpacs,preprintnumbers,amsmath,amssymb,footinbib,superscriptaddress,showkeys,floatfix]{revtex4-2}

\usepackage[english]{babel}
\usepackage[T1]{fontenc}
\usepackage{xcolor}
\usepackage[breaklinks=true,colorlinks,citecolor=blue,linkcolor=blue,urlcolor=blue]{hyperref}
\usepackage{graphicx}
\usepackage{amsmath}
\usepackage{amsfonts}

\newcommand{\vsaw}{v_{\mathrm{SAW}}}

\begin{document}

\title{Electron transfer between surface-acoustic-wave-induced \\ moving and static quantum dots}
\author{Mikel Olano}
\email{mikel.olano@dipc.org}
\affiliation{Donostia International Physics Center (DIPC), E-20018, Donostia-San Sebasti\'{a}n, Spain}
\affiliation{Polymers and Advanced Materials Department: Physics, Chemistry, and Technology, University of the Basque Country (UPV/EHU), E-20018, Donostia-San Sebasti\'{a}n, Spain}
\author{Geza Giedke}
\affiliation{Donostia International Physics Center (DIPC), E-20018, Donostia-San Sebasti\'{a}n, Spain}
\affiliation{Ikerbasque, Basque Foundation for Science, E-48013, Bilbao, Spain}
\date{\today}



\begin{abstract}
Fast long-range interactions between distant quantum dots in arrays remains an unsolved issue, which can be key to solve scalability issues in quantum simulation and computation processes, particularly related to the overhead associated with quantum error correction schemes. Furthermore, transport between static and moving quantum dots, relevant in surface acoustic wave induced experiments, has not been studied in detail. This article presents a paradigmatic model for picturing this process, where non-adiabatic terms driving a two-state transfer process are derived and discussed. Moreover, the main effects in the spin state of the electron and its effect on the transfer probability of the loading are analyzed including the most relevant interaction in semiconductor heterostructure induced 2 dimensional electron gases i.e. the Rashba-Dresselhaus terms.
\end{abstract}

\maketitle

\section{Introduction}

Semiconductor quantum dots (QDs) are one of the first proposed \cite{Loss1998,Burkard1999} and intensely pursued platforms to host and control spin qubits and perform quantum computations in arrays of exchange-coupled QDs \cite{Burkard2023}. While the experimental progress in this field is impressive, a well-known limitation is the short-ranged nature of the available interaction that only couples neighboring spins. A number of remedies including the use of microwave resonators \cite{Burkard2006} or floating gates \cite{Trifunovic2012} have been proposed to extend the coupling range. Another way to address that problem is to transport the spin qubits either by shuttling them using time-dependent electric gates \cite{Taylor2005}  or using potentials induced by surface acoustic waves \cite{Barnes2000} to create moving quantum dots within which spin qubits can be transported. All these approaches hold the promise of a stronger connectivity between qubits and a better scalability, reducing, in particular, the overhead associated with quantum error correcting schemes \cite{Pino2021,Sterk2022,Bluvstein2022,Bluvstein2023}.

High-fidelity shuttling over $\mu$m distances has been demonstrated experimentally both with gate-based driving in Si- \cite{Mills2019,Zwerver2023,Matsumoto2025} and with SAW-driven transport GaAs-based systems \cite{McNeil2011,Ford2017,Vandersypen2019,Zwerver2022,Wang2024}.
Since double quantum dot systems and their energy structure dependencies on multiple parameters have been thoroughly studied, concatenated tunneling through several neighboring QDs has been proposed as a transport method \cite{vanderWiel2002,Langrock2023,Buonacorsi2020}. In the simplest case, one would want to change the Hamiltonian such that the initial state follows an adiabatic trajectory and avoids populating unwanted states while changing the properties of the evolving one. The main drawback of these protocols lies in the time domain. Since the adiabatic approximation requires the process to be ``slow'' to avoid unwanted population exchanges caused by the Hamiltonian changing too fast, decoherence becomes a problem when dealing with the overall state's evolution. To overcome this, optimal control methods can be used \cite{Machnes2011,Mortensen2018,GuryOdelin2019}, which allow a faster change of the Hamiltonian that ``shorten'' the transfer processes by populating intermediate or final excited states by non-adiabatic transitions.

Recent improvements in the creation and control of SAWs have allowed to create single traveling minimum confined in a gate-defined channel \cite{Wang2022,Wang2024}. This allows the possibility of capturing a single electron in said minimum and making it interact with other electrons located in nearby static quantum dots, thus allowing for long-range interaction between several particles. Several different uses have been proposed (while some of them have been experimentally realized) for this set-up, such as single-qubit gates and measurement \cite{Doran2004,Bauerle2018}, two-qubit gates \cite{Barnes2000,Lepage2020}, and entanglement of distant electron spin qubits \cite{Benito2016,Jadot2021} or nuclear spins \cite{Bello2022}.

If this set-ups combining static dots to read and initialize spin states with ``moving'' quantum dots that transport the electrons around include a fast loading/unloading process, a non-adiabatic process is needed between the static and moving states since these are two different eigenstates of the complete Hamiltonian, which includes both static and moving potentials in the beginning/end of the process. Although some theoretical work has been done on the possible excitations that the electron transport induces in the particle's state once its loaded \cite{Takada2019}, the transfer process from being on a static qubit to a moving one has not been tackled yet. As precise as any loading process is, there is nothing ensuring that excitations have not already occurred at this stage. Moreover, the spin-orbit interaction that comes with symmetry breaking caused by the nanostructure holding the electron may introduce a spin state change that introduces an information loss that continues to propagate through the evolution of the system.

The purpose of this article is to study the electron transfer in a simple, paradigmatic setup and to present a model that covers the main interactions in the transfer process of a single electron in a static QD into a SAW minimum. Being GaAs heterostructures of high interest for their piezoelectric properties, the main interactions that affect the spin state of the moving quantum dot, i.e., the Rashba-Dresselhaus spin orbit interaction \cite{Huang2019}.

\section{System} 

We consider the dynamics of a single electron in a quantum well in a time-dependent trapping potential, that is composed of a static component modeling a quantum dot and a time-dependent one, describing the ``moving quantum dot'' created by a surface-acoustic-wave (SAW) pulse. As in the experiments \cite{McNeil2011,Bertrand2016} the pulse propagates in a 1D wave guide which confines the SAW laterally.

\begin{figure}[h!]
    \centering
    \includegraphics[width=0.7\linewidth]{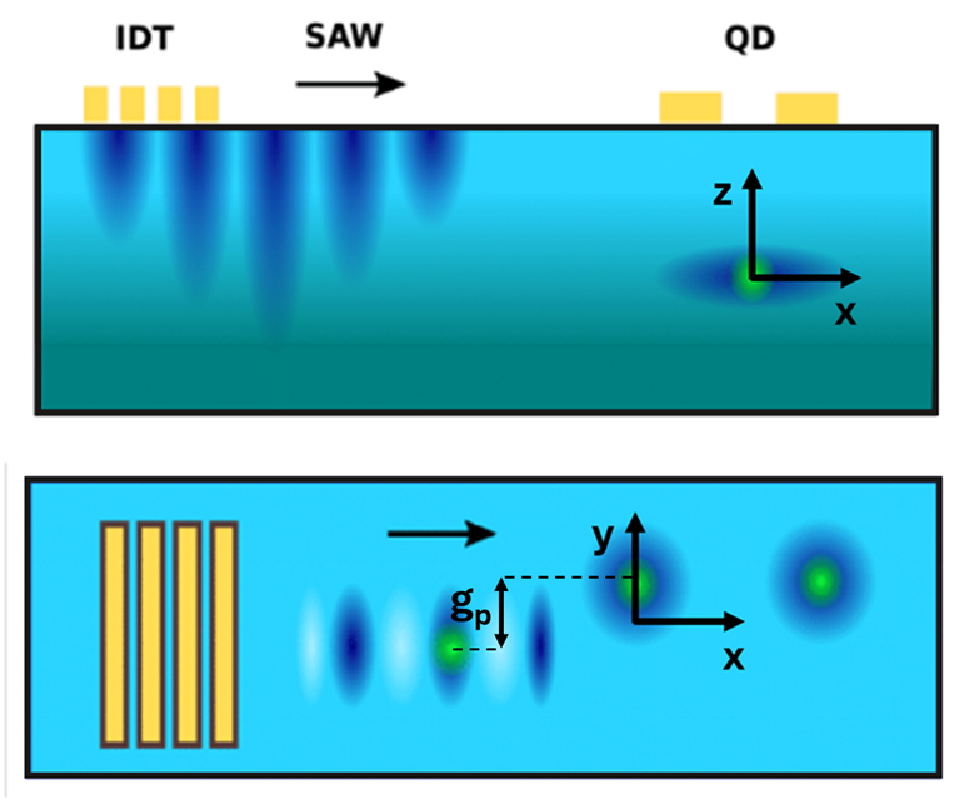}
    \caption{ Sketch showing the main characteristics of the
      system. The top figure shows a lateral view, where the electron
      is confined in the 2DEG by the top gates defining the QD and the
      SAW created by the IDT into penetrates roughly one wavelength
      into the substrate. The bottom figure shows a top view and the
      time-dependent lateral distance $d$
      between the global minima of the two potentials.}
    \label{fig:fig1}
\end{figure}

We are interested in transferring the electron from the moving QD into the static one and retrieving it from there and in the conditions required for these processes to happen with high reliability. We model our electrons by effective mass particles moving in the 2D quantum well, tracing out the perpendicular degree of freedom as in this strongly confined direction no motion is possible.

Thus we investigate the electron motion generated by the following 2D single-particle Hamiltonian

\begin{equation}
    H_0(t) = \frac{\pmb{p}^2}{2\; m^*} + V_s(\pmb{r}-\pmb{r}_s)+V_m(\pmb{r}-\pmb{r}_m(t)),
    \label{eq:free_H}
\end{equation}

\noindent where both $V_s(\pmb{r})$ and $V_m(\pmb{r})$ are potentials with a single global minimum at $\pmb{r}=0$ and the time-dependence only arises from the uniform motion of $\pmb{r}_m(t) = (x_m(0)+v_\mathrm{SAW}t,y_m(0))$ of the SAW-induced potential. For the time being we neglect spin-orbit coupling, which is weak in GaAs and consider its effect in Section \ref{sec:soi} below. We denote the time evolution generated by $H_0(t)$ by

\begin{equation} \label{eq:H0_evol}
    U_0(T,0) = \mathcal{T}\exp\Big[\frac{-i}{\hbar}\int_{0}^{T}ds H_{0}(s)\Big],
\end{equation}

\noindent where $\mathcal{T}$ denotes time ordering. 

Primarily, we are interested in the transfer probability

\begin{equation}\label{eq:PT}
    P_0(\Psi_i,\Psi_t) = |\langle \Psi_t|U_0(T,0)|\Psi_i\rangle|^2,
\end{equation}

\noindent from the initial state $|\Psi(0)\rangle =|\Psi_i\rangle$ localized at time $t=0$ in the static potential (which is well separated from the moving one at that time) to the target state $|\Psi_t\rangle$ localized in the moving potential at a time $t=T$ when the two minima are again far away from each other and how it depends on the choice of parameters of the two potentials.

Specifically, we consider initial configurations in which at $t=0$ the two potential minima are located at $x_s=0$ and $x_m=-L/2$, respectively and a SAW pulse traveling in $x$-direction with fixed speed $v_\mathrm{saw}$ such that at time $t=T$ the moving minimum is at $x_m(T)=L$. At this points, for the transfer process ending in the moving dot, one must be aware of the effect that the moving dot has in its groundstate. The equivalent of the energy eigenstates for the moving potential can be easily obtained: in the potential's rest frame, these are just its regular eigenstates. Hence, back in the lab frame the eigenstate displaced by $p_0=m^*\vsaw$ in momentum and $x_m=\vsaw t$ in position describes the appropriate wave function. The transverse separation of the two minima is denoted by $g_p = |y_m(0)-y_s|$ (see Figure \ref{fig:fig1}).  Typical QD sizes in the reference experiments are on the order of $l_{QD}\sim 10-100\; \textrm{nm}$ and usual SAW velocity units are $\vsaw\sim \textrm{nm}\cdot\textrm{ps}^{-1}$, which sets a range for the total process time to occur between $T\sim 10-100 \;\textrm{ps}$. 

In pioneering work on SAW-driven transfer of single electrons pulsed gates were used to inject and eject the electron \cite{McNeil2011,Bertrand2016,Takada2019,Wang2024}. We want to identify the conditions for transfer solely driven by the SAW pulse without any additional time-dependence. This will make it advantageous to work with potentials whose ground states (between which we want to transfer the electron) are close to degenerate. This can be ensured by choosing potentials whose harmonic approximation around their minimum have similar values for both the minimum value $V_0$ and the potential's frequencies $\omega_x,\omega_y$. Our standard choice will be a rotationally symmetric Gaussian potential $V_u(\pmb{r})=-V_{u,0}\exp(-k\|\pmb{r}\|^2)$ with $V_{u,0}>0$.

We will compute $U(T,0)$ by numerical integration. Before that, let us note that, as we shall see, the dynamics we are
interested in is in general non-adiabatic, as due to the near degeneracy of the two potentials, different instantaneous eigenstates are coupled. However,
transitions between non quasi-degenerate eigenstates will be seen to be negligible so that the relevant dynamics takes place in a two-dimensional subspace.

\section{Transfer process}

We now turn to the transfer probability as a function of different geometrical parameters characterizing the trapping potentials.  

For the calculations, we choose system parameters following values found in the relevant literature for GaAs experiments \cite{Tarucha1996,Stotz2005,Huang2019,Lepage2020}. The SAW speed has been chosen to be $\vsaw = 3\;\textrm{nm}\cdot\textrm{ps}^{-1}$ and for the Gaussian trapping potential, we fix the minimum value of the static dot to  $V_{s,0}=45\; \textrm{meV}$. To define the approximation around it with the usual harmonic potential's terms $k_{s}=m^*{\omega_0}^2/2V_{s,0} =3.96\;\textrm{nm}^{-2}$ such that the energy gap above each trap's ground state is $\hbar\omega_0=3 \; \textrm{meV}$. The minimum value difference between the potential minima ($V_{s,0}-V_{m,0}$) that maximizes the transfer probability for all potentials corresponds to the kinetic energy of the moving potential's groundstate, $V_{s,0}-V_{m,0}=m^*\vsaw^2/2$, which makes both potentials' groundstates resonant at large distances. To maintain the trapping frequency equal, we change $k_m = k_s\cdot V_{s,0}/V_{m,0}$.

Apart from the reference Gaussian function, other two potentials of different shape that share the harmonic approximation around their minimum with the reference potential have been used. This way, the near-degenerate subspace is maintained while varying the functional form of the potential to see to what extent common effects hold. The chosen potentials are a squared cosine function truncated for a single period and a Gaussian in the $y$ direction multiplied by the squared cosine truncated with a hyperbolic tangent difference in $x$. The explicit form of these potentials such that they share the same harmonic approximation around their minimum are

\begin{align}
    \begin{split}
    V_{\mathrm{Gauss}} &= -V_0 \exp(-k||\mathbf{r}||^2) \\
    \end{split}\label{eq:V_gauss}
\end{align}
\begin{align}
    \begin{split}
    V_{\mathrm{cos}}  = -V_0&\cos^2(\sqrt{k}x)\cos^2(\sqrt{k}y)\cdot \\ &\cdot\mathbf{H}(\pi/2-\sqrt{k}x)\mathbf{H}(\pi/2-\sqrt{k}y) \\
    \end{split} \label{eq:V_cosqr}
\end{align}
\begin{align}
    \begin{split}
    V_{\mathrm{Train}} &= -V_0\exp(-ky^2)\cos^2(\sqrt{k}x)\cdot \\ &\cdot[\tanh(x+\pi/2\sqrt{k})-\tanh(x-\pi/2\sqrt{k})]/2 
    \end{split} \label{eq:V_train}
\end{align}

\noindent where $\mathbf{H}$ is the Heaviside function and the truncation done by the hyperbolic tangent leaves a single absolute minimum with small minima to its side.

\begin{figure}[h!]
    \centering
    \includegraphics[width=\linewidth]{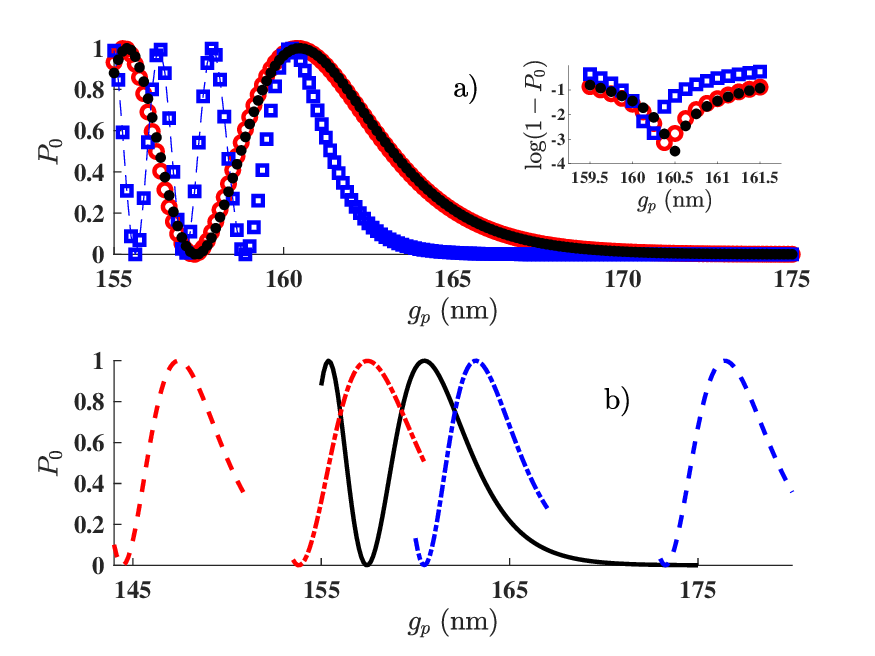}
    \caption{ Probability \eqref{eq:PT} to transfer the spinless particle's state from the static to the moving QD with: a) different potentials as a function of $g_p$ and for three potential shapes: Gaussian (black dots) \eqref{eq:V_gauss}, squared cosine (blue squares) \eqref{eq:V_cosqr}, and the combination squared-cosine in $x$, Gaussian in $y$-direction (red circles) \eqref{eq:V_train}. The inset is a closeup of the rightmost transfer probability peak, showing the logarithm of $1-P_0$ to see how close each peak is to unity; b) The probability curve near $g_{p0}$ for changed parameters: $\omega = 1.1\;\omega_0$ (dashed red), $v=2\;\vsaw$ (dot dashed red), $v=\vsaw /2$ (dot dashed blue) and $\omega = 0.9\;\omega_0$ (dashed blue)}
    \label{fig:TPvsgp}
\end{figure}

With the other parameters fixed, we first consider the influence of the impact parameter $g_p$ on the transfer probability \eqref{eq:PT}. For the integration of the time-dependent  Schrödinger equation, we use the simplest Trotter-Suzuki decomposition \cite{Suzuki1976,Hatano2005} to separate position- and momentum-based operators in \eqref{eq:H0_evol} up to a small correction. The Fourier transform can be used for an efficient implementation of both Trotter steps ensuring diagonality of the generator in each step. Figure \ref{fig:TPvsgp}a shows the transfer probability as a function of $g_p$. We focus on the larger possible values for $g_p$ for which transfer probabilities are closer to one, called $g_{p0}$ from now on. The typical behavior is that displayed by the black dots in Fig.~\ref{fig:TPvsgp}a: for large distances, no transfer is effected because the coupling between the two minima is too weak. For smaller $g_{p0}=160.5 \; \textrm{nm}$ a transfer probability of $P_0 = 0.9997$  is achieved, while further reduction of $g_p$ leads to an oscillation of $P_0$ between 0 and 1 on a scale of a few nanometers. 
Each symbol in the figure represents the result of the  complete evolution from $t=0$ to $t=T$. In view of the continuity of $P_0$ as a function of $g_p$, further similar figures will show a line.

The almost perfect overlap between the black and red points, together with the dissimilarity between them and the blue points, suggests that the potential's lateral extension is very determining in this transfer process. In addition to this, the rightmost peak occurs for the same value of $g_p$ for the three relevant potentials, suggesting some common effect coming from the potential's strength at its minimum. The eigenstates of $H_0$ are not well defined by a linear combination between the individual potentials' grounstates, making it difficult to have an estimation of this peak related to handy parameters such as the width of the states or a tunneling time estimate.

The dependence of the transfer probability on other system parameters is showcased in Fig.~\ref{fig:TPvsgp}b: it only shows $P_0$ around $g_{p0}$ for different values of SAW speed and trapping frequency. Together with Fig.~\ref{fig:TPvsgp}a this shows that the position $g_{p0}$ of this optimum depends on system parameters such as potential transversal shape, its trapping frequency, or the speed of the moving potential, but that the peak value very close to $P_0=1$ is not strongly affected (as long as the resonance condition is ensured). Note that $g_{p0}$ is more sensitive to $\omega$ than to $\vsaw$: a ten percent change in the potential strengths causes the maximum to be shifted more than $10 \; \textrm{nm}$, whereas doubling or halving $\vsaw$ moves it around $\sim 3 \; \textrm{nm}$. This reinforces the importance of the lateral confinement of both potentials over other parameters in the transfer process.

\begin{figure}[ht]
    \centering
    \includegraphics[width=\linewidth]{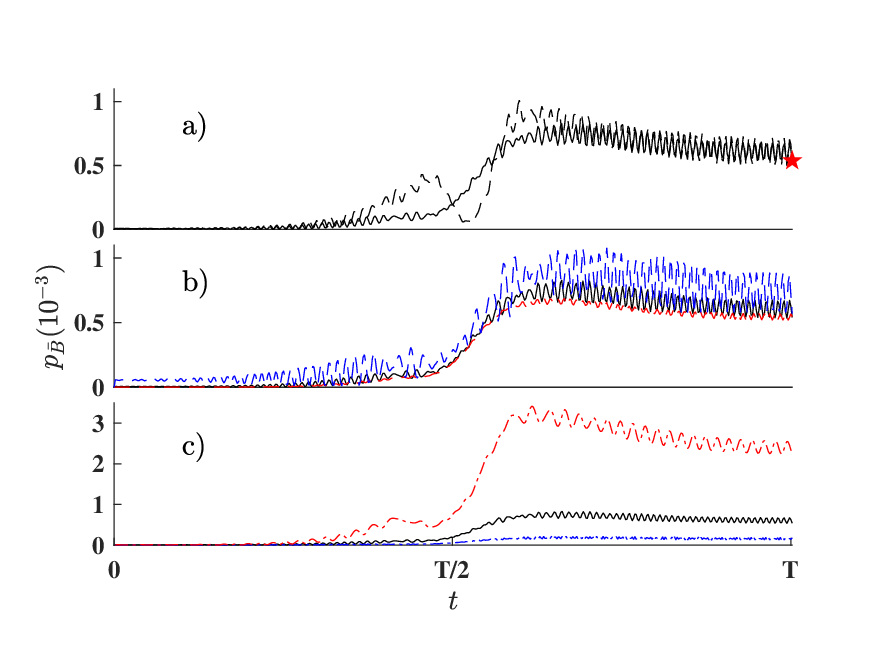}
    \caption{Probability of finding the state out of the two first instantaneous eigenstates $p_{\overline{B}}(t)$ during its evolution for $g_{p0}=160.5\;\textrm{nm}$ (continuous black) is shown as a reference in all figures. The red star shows the probability of finding the first excited state of the moving dot in the evolved state at $t=T$. The rest of the curves are $p_{\overline{B}}(t)$ for: a) $g_{p1}=155.375\;\textrm{nm}$ (dashed black), the leftmost peak in the Gaussian transfer probability curve in \ref{fig:TPvsgp}; b) $g_{p0}=176.375\;\textrm{nm}$ and $\omega = 0.9\;\omega_0$ (dashed blue) and $g_{p0}=147.375\;\textrm{nm}$ and $\omega = 1.1\;\omega_0$ (dashed red); c) $g_{p0}=157.5\;\textrm{nm}$ and $v=2\;\vsaw$ (dot dashed red) and $g_{p0}=163.25\;\textrm{nm}$ and $v=\vsaw /2$ (dot dashed blue).}
    \label{fig:exc_probs}
\end{figure}

Knowing already the $g_p$ values for various high-fidelity points, one can ask whether the excitation at the maximum transfer probability occurs only in the two-fold groundstate $B=\{|E_0\rangle,|E_1\rangle\}$ or if some part of the evolved state excites further. The probability of finding the state in the two first instantaneous eigenstates $p_B$ of the system can be calculated at various points along its path. The remaining probability $p_{\overline{B}} = 1-p_B$ belongs to the proportion of the state that has been excited.
The time dependence of this probability is shown in figure \ref{fig:exc_probs} for the different curve maxima shown in figure \ref{fig:TPvsgp}b (the precise number are mentioned in the caption). All of them show two common characteristics: their increase around the middle of the process and the oscillating nature from there on. The former suggests that the excitation processes occur at the same time of the delivery of the particle to the moving dot, while the latter comes mostly from the phase difference between the instantaneous eigenstates out of $B$ that have been populated and the proportion of $B$ eigenstates in the evolved state. This can be confirmed by looking at two things: the probability of finding the first excited state of the moving potential in the evolved state at time $T$ $P_0(E_0,E_{m,1})=0.54\cdot10^{-3}$, which is marked by a red star in figure \ref{fig:exc_probs}a; and the frequency of the oscillations in the second half $\hbar\omega_{P}=2.8\; \textrm{meV}$. Knowing that the moving potential's groundstate at the end of the process can be described as the instantaneous groundstate of the moving potential in a moving frame ($|\Tilde{E}_{m,0}(T)\rangle = \exp(ip_0\hat{x}/\hbar)|E_{m,0}(T)\rangle$ with $p_0 = m^*\vsaw$ where $(T+V_m(T))|E_{m,0}(T)\rangle = E_{m,0}|E_{m,0}(T)\rangle$) one can see that the majority of the mean value of $p_{\overline{B}}(T)$ comes from not including the first excited state of the moving potential in $B$, since $|\langle E_{m,1}(T)|\exp(ip_0\hat{x}/\hbar)|E_{m,0}(T)\rangle|^2= 0.59\cdot 10^{-3}$. According to the figures, the parameter that mostly varies the proportion of the state that belongs to the excited states at the end of the process is the SAW's speed, since it is the only parameter that, changed, increases the probability such that $p_{\overline{B}}(T)>10^{-3}$. Even then, one has to take into account that there is a higher loss of probability to find the final state in $B$ when the speed of the SAW is increased.

\begin{figure}[ht]
    \centering
    \includegraphics[width=\linewidth]{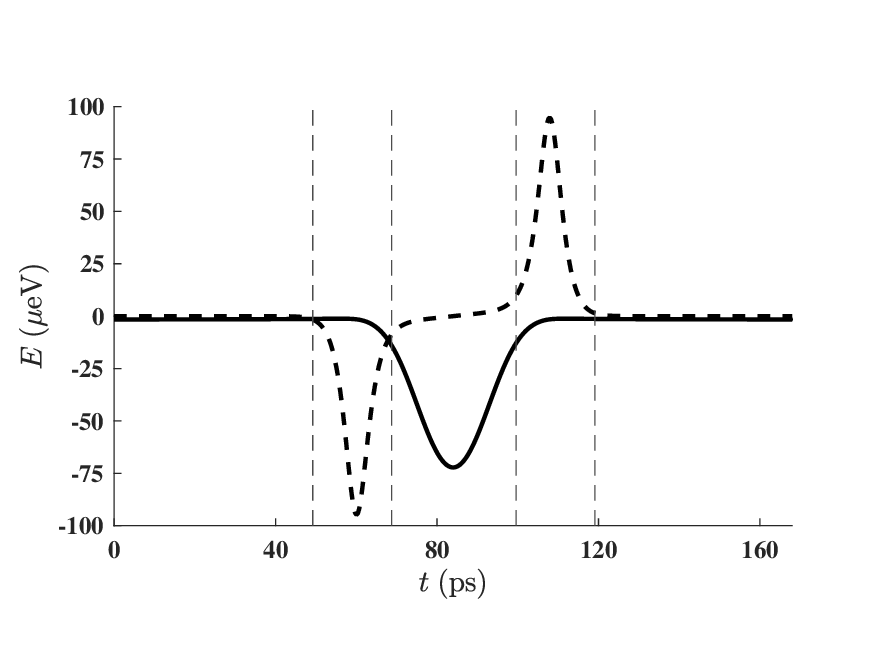}
    \caption{Time-dependent functions that enter in the few-level description of the spinless electron's evolution for $g_{p0}=160.5\;\textrm{nm}$ and $\omega_0 = 3\;\textrm{meV}$: $f_z$ (solid) and $f_y$ (dashed). The vertical dashed-lines define $t_0$, $t_1$, $t_2=T-(t_0+t_1)$ and $t_3=T-t_0$, respectively.}
    \label{fig:evolterms}
\end{figure}

Whether the error comes from the SAW's speed or the transfer process populating the excited states, knowing that the probability of the evolving state not to be described by the $B$ basis is as low as $p_{\overline{B}}(t)<10^{-3}$ for all $t\in [0,T]$ allows us to look only at the Hamiltonian terms in this subspace to describe the transfer process.

\section{The effective model}

Once the eigenstates and energies have been calculated along the path, one should tackle the adiabaticity of the process. Since the unperturbed Hamiltonian is time dependent, so are its eigenstates, which inserted in the Schr\"{o}dinger equation give a time dependent coupling between eigenstates $|E_j(t)\rangle$ and $|E_k(t)\rangle$ of the form

\begin{equation} \label{eq:Omega_jk}
    \Omega_{jk} = -i\hbar\langle E_j|\Dot{E}_k\rangle = -i\hbar\frac{\langle E_j|\Dot{H}_0|E_k\rangle}{E_k - E_j}.
\end{equation}

Note that, due to the components of \eqref{eq:free_H} being all real, one can choose the eigenstates to be real at all times, which we do for our calculations. This leaves us with an effective Hamiltonian $\Tilde{H}_0$ that includes real values in the diagonal (the energies of the system) and a purely imaginary coupling between the two states. In terms of Pauli matrices, one can write such a matrix as

\begin{align}
\begin{split}\label{eq:2dHam}
    \Tilde{H}_{0}(t) &= \frac{\varepsilon_0 + \varepsilon_1}{2}\sigma_0 +  i\Omega_{01}\sigma_y + \frac{\varepsilon_0 - \varepsilon_1}{2}\sigma_z =  \\ &= f_0(t)\sigma_0 + f_y(t)\sigma_y + f_z(t)\sigma_z.
\end{split}
\end{align}

Figure \ref{fig:evolterms} shows the time-evolution of the last two terms. The first one, whilst necessary to determine the relative phase between the remaining part of the state in the static dot with respect to the moving one, is not too relevant when the transfer process is optimized as it is here. First, there is an obvious symmetry to the parameters characterizing this effective Hamiltonian, that is related to a symmetry in the set-up of our problem: the potentials we consider have a reflection symmetry ($V_{u}(-x,y)=V_u(x,y)$ for both $u=s,m$) and therefor the Hamiltonian $H_{0}(t)$ satisfies 
\begin{equation}
    H_{0} = R_x H_{0}(t)R_x^\dag,
\end{equation}
where $t\in[0,T]$ and $R_x$ is the unitary implementing reflection at $x=0$. This, in turn, means that the eigenstates are also symmetric with this reflection when inverting time $|E_n(t)\rangle = R_x|E_n(T-t)\rangle$, as well as the energies (and, therefore, their differences) $E_n(t)=E_n(T-t)$. Looking at \eqref{eq:Omega_jk}, one can also confirm the antisymmetry of $f_y$, since

\begin{align}
\begin{split}
  &\Omega_{jk}(t) = \frac{\langle{E_{j}(t)}|\frac{d}{dt}H_0(t)|{E_{k}(t)}\rangle}{E_{k}(t)-E_{j}(t)} =\\
  =&\frac{\langle{E_{j}(T-t)} | (-1) H_0(T-t)|{E_{k}(T-t)}\rangle}{E_{k}(T-t)-E_{j}(T-t)}=\\
  =&-\Omega_{jk}(T-t).
\end{split}
\end{align}

Figure \ref{fig:evolterms} shows three different zones, two of them dominated by $f_y$ (which breaks the adiabaticity condition) and the center one where $f_z$ is stronger. The outer vertical lines mark the moment in which $|f_y/f_z|>10$, whereas the inner ones are the mean values of the points where $|f_y/f_z|<10$ and $|f_z/f_y|>10$. Let us name these points $t_j$ for $j=\{0,1,2,3\}$ in increasing order. Considering the dominance of each one of the functions in the different time periods, one can define each partial unitary by their appearing order $j$ and dominant term $i\in\{y,z\}$ as

\begin{equation}
    U_{j,i} = \mathcal{T}\exp\Big(\frac{-i}{\hbar}\int^{t_{j+1}}_{t_j} f_i(t')dt'\Big)
\end{equation}

\noindent so that the unitary that evolves the electron can be approximated by neglecting the smaller terms in each interval

\begin{align}
\begin{split}
    U_0(t_3,t_0) &\simeq U_{2,y}U_{1,z}U_{0,y}.
\end{split}
\end{align}

Taking into account the symmetry considerations in the previous section, the exponent in the first and last applications of the partial unitaries have the opposite sign, making the corresponding time evolutions inverse to each other. One can also infer that $t_3 = T-t_0$ and $t_1-t_0=t_3-t_2$. This sets a series of conditions for a complete transfer between the two eigenstates under this approximation. We define 

\begin{align}
    \gamma=\frac{1}{\hbar}\int^{t_1}_{t_0}f_y(t')&dt' = -\frac{1}{\hbar}\int^{t_3}_{t_2}f_y(t')dt'   \label{eq:gamma}\\
    \Gamma = & \frac{1}{\hbar}\int^{t_2}_{t_1}f_z(t')dt'.  \label{eq:Gamma}
\end{align}

Using the expression of the exponentiation of a general combination of Pauli matrices $\exp(i\theta\;\vec{v}\cdot\vec{\sigma}) = \cos\theta \cdot\mathbb{I}+ i\sin\theta\cdot \vec{v}\cdot\vec{\sigma}$, one can rewrite the approximated unitary as

\begin{multline} \label{eq:U_app}
U_0(t_3,t_0) \simeq \left(
  \begin{matrix}
    \cos \Gamma +i\sin \Gamma \cdot \cos 2\gamma \\ 
    -i\sin\Gamma\cdot\sin 2\gamma  \\ 
  \end{matrix}\right.                
\\
  \left.
  \begin{matrix} 
    -i\sin\Gamma\cdot\sin 2\gamma \\ 
    i\sin\Gamma\cos 2\gamma
  \end{matrix}\right).
\end{multline}

This sets a series of conditions for a complete transfer between the initial and target instantaneous eigenstates. For a complete transfer between them to be accomplished the following conditions should hold:

\begin{equation} \label{eq:cond1}
    \Gamma = \frac{\pi}{2}+n_{\Gamma}\pi
\end{equation}

\noindent and

\begin{equation} \label{eq:cond2}
    \gamma = \frac{\pi}{4}+n_{\gamma}\frac{\pi}{2}.
\end{equation}

\noindent where $n_\gamma,n_\Gamma\in \mathbb{Z}$ are integer numbers for which the final evolution gives a complete state transfer and becomes the $X$ gate up to a global phase. Calculating the value of both parameters for the range of $g_p$ shown in figure \ref{fig:TPvsgp} one sees that $\gamma$ is close to constant with a value of $-\pi/4$. The $U(t_3,t_2)$ and $U(t_1,t_0)$ then become basis change matrices for even and odd combinations of the original states. This simplifies the unitary in \eqref{eq:U_app} to $U(t_3,t_0) = e^{i\Gamma\sigma_x}$, which gives a probability of $P_{T,0}(T)=\sin^2\Gamma$ to find the particle in the other potential at the end of the process. Also, since the transfer process can be simplified to this two-level system as a good approximation, the insertion of the electron from the initial moving state to a final static state would be as optimal and require the same conditions to be fulfilled.

\section{Introducing the spin-orbit interaction}\label{sec:soi}

The unperturbed Hamiltonian in equation \eqref{eq:free_H} does not take into account the spin degree of freedom. However, in GaAs the presence of structural inversion asymmetry and bulk inversion asymmetry give rise to both Rashba \cite{Rashba1984} and Dresselhaus \cite{Dresselhaus1955} spin orbit coupling. Denoting by $x$ and $y$ the main crystallographic directions in the (001) plane of GaAs, 
the Hamiltonian takes the form

\begin{align}
\begin{split} \label{eq:H_SO}
    H_{SO} = \alpha_R(\hat{p}_x&\sigma_y-\hat{p}_y\sigma_x) + \\ &+ \beta_D(-\hat{p}_x\sigma_x+\hat{p}_y\sigma_y). 
\end{split}
\end{align}

The perturbative nature can be already intuited from the size of $\alpha_{R}$ and $\beta_D$, which in the case of both GaAs and Si lie on the range $\sim 0.1\;\mu \textrm{m}\cdot\textrm{ns}^{-1}$ \cite{Studer2009,Dettwiler2017}. Considering the momentum of the moving potential's groundstate as a mean value for $p_0 = \langle p_x\rangle$, the expression of \eqref{eq:H_SO} gives a interaction size range of $\sim m^*\vsaw^2/10\ll \hbar\omega_0$, which is five times smaller than an already small quantity in the energy range of $H_0$, the kinetic energy of the moving particle, which in turn is much smaller than the energy gap between the groundstates and the excited states. Following \cite{Huang2019}, we can define new in-plane axes $x',y'r$ rotated by $\pi/4$ so that $p_{x'}=(p_x+p_y)/\sqrt{2}$ and $p_{y'}=(p_x-p_y)/\sqrt{2}$ and similar for $\sigma_{x'},\sigma_{y'}$. In this basis, which we will from now on adopt, dropping the prime to simplify the notation, $H_{SO}$ takes the simpler form

\begin{equation}\label{eq:Hsoi2}
    H_{SO} = (\beta_D-\alpha_R)\hat{p}_y\sigma_x+(\beta_D+\alpha_R)\hat{p}_x\sigma_y.
\end{equation}

From this it is evident that the cases $\alpha_R=\pm\beta_D$ lead to a particularly simple evolution as only one spin operator remains and thus turns into a conserved quantity. We consider these two cases for the value $\alpha_R = 0.3\;\textrm{nm}\cdot \textrm{ps}^{-1}$ and  project $H_{SO}$ to the time-dependent two-dimensional orbital subspace studied in the previous section to study its effect on the transfer of a spin qubit.

\begin{figure}[ht]
    \centering
    \includegraphics[width=\linewidth]{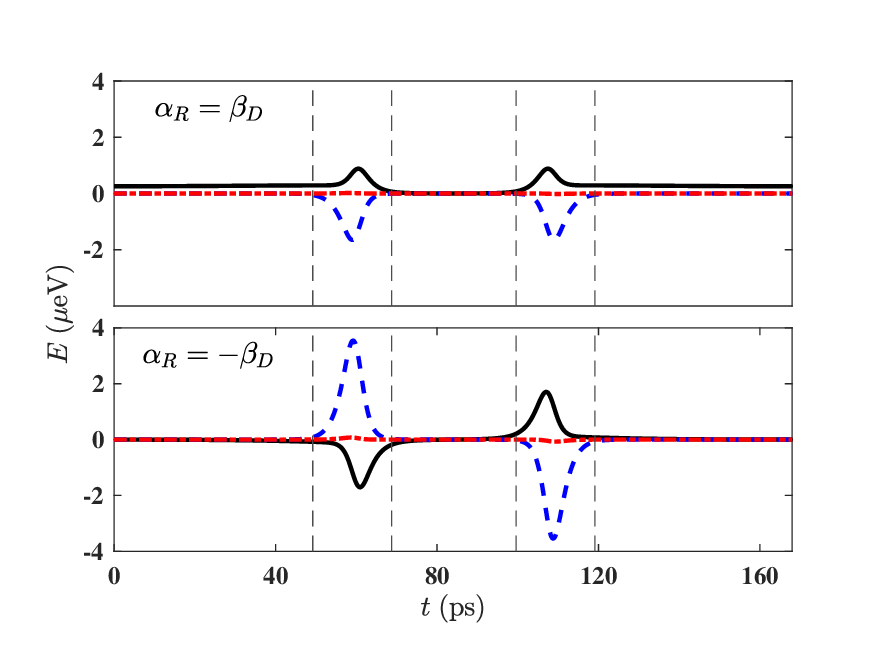}
    \caption{Time-dependent correction functions that enter in the few-level description of the electron's evolution due to spin-orbit interaction. The corrections are $\delta f_z$ (solid black), $\delta f_y$ (dotted-dashed red) and $\delta f_x$ (dashed blue) for the two Rashba and Dresselhaus parameter configurations that we are interested in. Parameters (and consequently $H_0$) are the same as in \ref{fig:evolterms}, as well as the vertical lines.}
    \label{fig:evolterms_2}
\end{figure}

The unperturbed Hamiltonian is spin-independent and thus can be written as a direct sum, the same $2\times2$-matrix as given in \eqref{eq:2dHam} acting on the spin-up and on the spin-down component of the electron. As before, we discard the constant energy shift of the subspace. To this we now add the perturbation, which respects the block structure (not mixing different spins) but the sign of the perturbation differs for the two spin components, leading to a total block-diagonal $4\times4$ Hamiltonian of the form

\begin{equation}
    H(t) = H_+(t)\oplus H_-(t)
\end{equation}

with 

\begin{equation} \label{eq:parameterized_evol_pm}
    H_\pm(t) = [f_z\pm\delta f_z]\sigma_z + [f_y\pm\delta f_y]\sigma_y \pm \delta f_x\sigma_x,
\end{equation}

where all the functions $f_\alpha,\delta f_\alpha$ are time-dependent and where the signs refer to the spin quantum number. This can be explained taking into account that, unlike the off-diagonal terms coming from the unperturbed Hamiltonian, the spin-orbit interaction is symmetric with respect to time-reversal.  Looking at figure \ref{fig:evolterms_2}, one can see first,  that the terms are in fact perturbative compared to the orbital terms depicted in Fig.~\ref{fig:evolterms}; and second, that the $\sigma_y$ term is much smaller than the other two and can be neglected even compared to the other two spin-orbit corrections. 

For the remaining terms, the same five time intervals as for $H_0(t)$ can be distinguished. For $\alpha_R=\beta_D$ the $\vsaw$ is along the momentum direction that appears in \eqref{eq:Hsoi2} leading to the constant component multiplying $\sigma_z$ in the first and last intervals in the upper panel of Fig.~\ref{fig:evolterms_2},  which in this treatment appears as a diagonal correction coming from the time derivative of the eigenstates $\Omega_{0\pm0\pm}$. This happens because the instantaneous eigenstates do not capture the movement of the particle, while their time-derivative does. Moreover, taking into account that the vertical lines are the same as in figure \ref{fig:evolterms}, the corrections to the unitary need to be done only for the time intervals $[t_0,t_1]$ and $[t_2, t_3]$. The correction to the complete unitary $U(t_3,t_0)$, though, will also depend on the unperturbed part $U_{0}(t_2,t_1) = e^{i\Gamma\sigma_z}$.

\begin{figure}[ht]
    \centering
    \includegraphics[width=\linewidth]{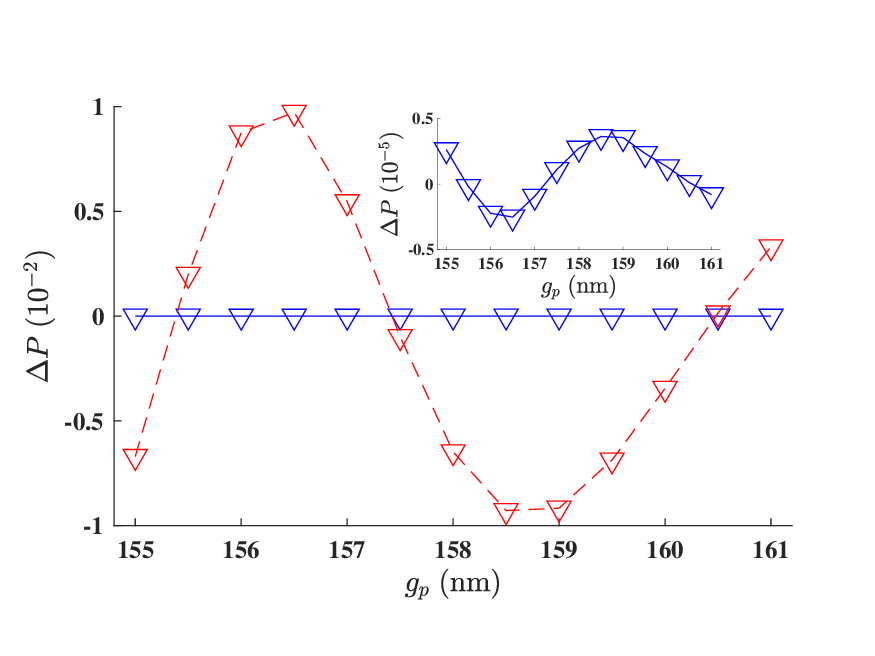}
    \caption{ Transfer probability difference $\Delta P$ caused by the inclusion of the spin-orbit interaction for $\alpha_R = \beta_D$ (dashed red) and $\alpha_R = -\beta_D$ (solid blue) for different values of $g_p$ in the case of the Gaussian potentials. The inset shows a scaled version of the latter, which can not be appreciated in the bigger picture.}
    \label{fig:deltaTPvsgp}
\end{figure}

Once these terms are calculated by diagonalizing the total Hamiltonian, one can make the assumption that the excited states of the system out of these four lowest energy eigenstates will not be populated since essentially the unperturbed Hamiltonians energy spectrum remains the same. The Trotterized approximation can be now used to evolve the same initial orbital state with a particular spin direction, now in this small subspace.
The effect of the perturbative terms on the probability to find the electron in the other potential can be seen in figure \ref{fig:deltaTPvsgp}, but only for the case where the initial state is in the positive spin direction. If the same state is initialized in the opposite spin direction, $\Delta P$ changes sign, which is an effect that can be traced back to the correction term signs in \eqref{eq:parameterized_evol_pm}. Moreover, its magnitude varies by three orders of magnitude going from the $\alpha_R = \beta_D$ to the $\alpha_R = -\beta_D$ (see inset), even though the latter is the case with the somewhat larger perturbation terms (see Fig.~\ref{fig:evolterms_2}). Looking at the points where this correction is minimal in absolute value, one finds that it happens at the same points where the transfer probability is extremal in figure \ref{fig:TPvsgp}: $g_{p1}$, the minimum at $g_p=157.375\;\textrm{nm}$ and $g_{p0}$. To explain these observations, we start by approximating the perturbed unitaries for the three different time intervals $U_j\equiv U(t_{j},t_{j-1}), j=1,2,3$ by their first order Dyson series

\begin{align}
\begin{split}
    U(t,t')\approx \; &U_{0}(t,t')- \\ &-i\hbar \int_{t'}^{t}ds
U_{0}(s,t)H_{SO}(s)U_{0}(t',s)
\end{split}
\end{align}

\noindent and, following previously defined forms of the  $U_j$ in the corrected time frames, one can follow by inserting $U_0(s,t) = \cos\gamma_{s,t}\mathbb{I}-i\hbar\sin\gamma_{s,t}\sigma_y$,
  
\begin{align}
\begin{split}
  U(t_{2},t_{1}) =\; & U_{0}(t_{2},t_{1}) - \\ & -i\hbar\int_{t_{1}}^{t_{2}}\big[\cos\gamma_{s,t_{2}}\cos\gamma_{t_{1},s}H_{SO}(s) -  \\ &  -\sin\gamma_{s,t_{2}}\sin\gamma_{t_{1},s} \sigma_{y}H_{SO}(s)\sigma_{y}-\\
  & -i\cos\gamma_{s,t_{2}}\sin\gamma_{t_{1},s} H_{SO}(s)\sigma_{y}- \\ & -i\sin\gamma_{s,t_{2}}\cos\gamma_{t_{1},s} \sigma_{y}H_{SO}(s)\big]ds.
\end{split}
\end{align}

\noindent where $\gamma_{t_1,s}+\gamma_{s,t_2}=\gamma=-\pi/4$. Since the spin-orbit correction has no $y$-component, $H_1(s)$ anticommutes with $\sigma_y$ and we can write the whole integrand as a linear combination of $\sigma_x$ and $\sigma_z$:

\begin{align}
\begin{split}
  U(t_{2},t_{1}) &= U_{0}(t_{2},t_{1})-\\ &-i\hbar\int_{t_{1}}^{t_{2}}\big[ \cos(\gamma_{s,t_{2}}-\gamma_{t_{1},s}) H_{SO}(s)+\\ &+ i\sigma_y \sin(\gamma_{s,t_{2}}-\gamma_{t_{1},s}) H_{SO}(s)\big]ds.\\
&\label{eq:3}\equiv U_{0}(t_{2},t_{1}) -i\hbar (K_x\sigma_x+K_z\sigma_z).\\
&\equiv U_{0}(t_{2},t_{1}) -i\hbar K.
\end{split}
\end{align}

A similar expression can be obtained for the interval $[t_2, t_3]$ (with primed functions to avoid confusion). Then the overall gate $U(t_3,t_0)$ is:

\begin{align}
\begin{split}
    U(t_3,t_0)\approx\; &U_0(t_3,t_0)-\\&-i\hbar\big[
    K' U^{(0)}_2U^{(0)}_1 +U^{(0)}_3U^{(0)}_2 K \big] = \\ &= U_0(t_3,t_0)+\Delta U_{SO},
\end{split}
\end{align}

\noindent where $U^{(0)}_j=U_0(t_{j},t_{j-1})$ and writing out the unperturbed unitaries in terms of their generators the correction given by the spin-orbit reads

\begin{align}
\begin{split}
    \Delta U=-i\hbar\big[ \cos\Gamma &(\cos\gamma K' -i\sin\gamma K'\sigma_y+ \\ &+\cos\gamma'
    K -i\sin\gamma'\sigma_y K)+\\
    +\sin\Gamma &(-i\cos\gamma K'\sigma_z+i\sin\gamma
      K'\sigma_x -\\ &-i\cos\gamma' \sigma_z K
      -i\sin\gamma'\sigma_x K)\big].
\end{split}
\end{align}

Following condition \eqref{eq:cond2} with $n_\gamma=-1$, as found by the previous analysis on the unperturbed Hamiltonian 

\begin{align}
\begin{split}\label{eq:DeltaU_final}
  \Delta U=-&\hbar\big[\cos\Gamma/\sqrt{2}(i\mathbb{I} +\sigma_y) (K'+K)+\\+\sin&\Gamma/\sqrt{2}(K'\sigma_z +\sigma_z K + K'\sigma_x+\sigma_x K)\big] =\\
  =-&\hbar\big[i\cos\Gamma(K_+^{(2)}\sigma_z + K_+^{(1)}\sigma_x) +\\&+\sin\Gamma(K_+^{(1)}\mathbb{I}-iK_-\sigma_y)\big]/\sqrt{2}
\end{split}
\end{align}

\noindent where

\begin{align}
    K_+^{(1)} &= K'_z + K_z + K'_x + K_x \\
    K_+^{(2)} &= K'_z + K_z - (K'_x + K_x) \\
    K_- &= K'_x - K_x + K_z - K'_z.
\end{align}

This is the expression for the first order correction of $U(t_3,t_0)$ due to the spin-orbit interaction for the unitary time evolution operator from $t_0$ to $t_3$. To see its effect on the transfer probability change, one can apply the total time evolution unitary to an initial state $|E_1(0)\rangle$ (which corresponds to the electron in the static dot at the beginning of the process) and look at the probability of finding the state that it has to be transferred to $|E_0(T)\rangle$:

\begin{align}
\begin{split}
    (U_0(T,0)+&\Delta U(T,0))|E_1(0)\rangle =\\ =&(U_{0,01}(T,0)+\Delta U_{01}(T,0))|E_0(T)\rangle + \\ &+  (U_{0,11}(T,0)+\Delta U_{11}(T,0))|E_1(T)\rangle
\end{split}
\end{align}

\noindent where the last to subscripts refer to the corresponding row and column of the element in the unitary time evolution's matrix, both for the unperturbed Hamiltonian $U_{0,ij}=\langle \Psi_i|U_0(T,0)|\Psi_j\rangle$ and the spin-orbit correction $\Delta U_{ij} = \langle\Psi_i|\Delta U|\Psi_j\rangle$. The probability is then

\begin{equation}
    P(E_1,E_0) = |U_{0,01}|^2+2\Re({U_{0,01}}^* \Delta U_{01})+|\Delta U_{01}|^2.
\end{equation}

Here, the first term corresponds to the transfer probability using the unperturbed Hamiltonian to evolve the chosen initial state, $P_0(E_1,E_0)=|U_{0,01}|^2$. The third term is a second order term in the correction, which can be neglected. It is the second term that shows the first order effect of the spin-orbit interaction on the transfer probability of the chosen initial state 

\begin{equation}
    \Delta P(E_1,E_0) = 2\Re({U_{0,01}}^* \Delta U_{01}) = \sqrt{2}\hbar\cos\Gamma\cdot K_+^{(1)}.
\end{equation}

Following the symmetry arguments as before, one can demonstrate that $K_+^{(1)}=0$ for the case $\alpha_R=-\beta_D$, which explains its much lower impact in the transfer probability change in figure \ref{fig:deltaTPvsgp}. When condition \eqref{eq:cond1} is met, $\cos\Gamma = 0$, which, in turn, means that there is first-order protection to the spin-orbit interaction perturbation for the probability transfer of the particle for both spin directions when the transfer process is optimal.

\section{Conclusions and outlook}

In conclusion, we have shown that under experimentally available parameters, close to complete transfer can be achieved between a static QD's groundstate and a moving dot's lowest-energy eigenstate simply by having the moving potential pass by the static QD in a suitable distance. Moreover, this transfer can be done involving only the two lowest eigenstates of the system including both potentials. For every parameter combination leading to an optimal transfer process, we have seen that the Hilbert's space reduction to these two almost degenerate eigenstates makes us lose a small part  of the dynamics that can mostly be attributed to transfer to the first excited state of the moving potential. While we have only considered transfer from the moving to the static potential, the reverse process is, in the highly symmetric situation we consider, simply obtained by transforming to the frame in which the SAW pulse is at rest, so that the same transfer probabilities apply.

After including the spin-orbit interaction and describing its effect on the effective Hamiltonian, we have seen that the maximum difference in transfer probability caused by this interaction is of the order of $10^{-2}$. However, from a first order Dyson series of the time-evolving unitary, we have derived an expression that shows a protection against spin-orbit interaction for the extreme values of the transfer probability curve.

Given the generality of many assumptions in this article, and taking into account the variety of optimal transfers this regime offers, we think that other QD setups may also benefit from this work, such as the dynamical quantum dot experiments \cite{Stotz2005} or other works that involve propagating potential minima with relative velocities \cite{Knrzer2018}.

As for the prospective work that this model leads to, there are two main paths to be followed: the inclusion of the time-evolving first excited eigenstate as part of the description of the system and the inclusion of two-particle interactions in order to formulate the possibly useful entangling gates that could be created by this proposal. The former is not only an attempt to improve the instantaneous Hilbert space along the process, but also a possibility of purposefully populating other excited states looking for STA schemes that improve the transfer probability's dependency on $g_p$, which we think could be the hardest experimental implementation feature. The latter would be a good step forward in the description of multipartite entanglement processes, as well as more interesting proposals towards quantum simulation experiments.

\begin{acknowledgments}
We acknowledge funding through the grant PID2023-146694NB-I00 funded by MICIU/AEI/10.13039/501100011033/ and by ERDF/EU, by the Basque Department of Education (PIBA-2023-1-0021), and by the European Union NextGenerationEU/PRTR-C17.I1 and  through the IKUR Strategy under the collaboration agreement between Ikerbasque Foundation and DIPC on behalf of the Department of Education of the Basque Government.
\end{acknowledgments}

\bibliography{bibliography}

\begin{thebibliography}{44}%
\makeatletter
\providecommand \@ifxundefined [1]{%
 \@ifx{#1\undefined}
}%
\providecommand \@ifnum [1]{%
 \ifnum #1\expandafter \@firstoftwo
 \else \expandafter \@secondoftwo
 \fi
}%
\providecommand \@ifx [1]{%
 \ifx #1\expandafter \@firstoftwo
 \else \expandafter \@secondoftwo
 \fi
}%
\providecommand \natexlab [1]{#1}%
\providecommand \enquote  [1]{``#1''}%
\providecommand \bibnamefont  [1]{#1}%
\providecommand \bibfnamefont [1]{#1}%
\providecommand \citenamefont [1]{#1}%
\providecommand \href@noop [0]{\@secondoftwo}%
\providecommand \href [0]{\begingroup \@sanitize@url \@href}%
\providecommand \@href[1]{\@@startlink{#1}\@@href}%
\providecommand \@@href[1]{\endgroup#1\@@endlink}%
\providecommand \@sanitize@url [0]{\catcode `\\12\catcode `\$12\catcode `\&12\catcode `\#12\catcode `\^12\catcode `\_12\catcode `\%12\relax}%
\providecommand \@@startlink[1]{}%
\providecommand \@@endlink[0]{}%
\providecommand \url  [0]{\begingroup\@sanitize@url \@url }%
\providecommand \@url [1]{\endgroup\@href {#1}{\urlprefix }}%
\providecommand \urlprefix  [0]{URL }%
\providecommand \Eprint [0]{\href }%
\providecommand \doibase [0]{https://doi.org/}%
\providecommand \selectlanguage [0]{\@gobble}%
\providecommand \bibinfo  [0]{\@secondoftwo}%
\providecommand \bibfield  [0]{\@secondoftwo}%
\providecommand \translation [1]{[#1]}%
\providecommand \BibitemOpen [0]{}%
\providecommand \bibitemStop [0]{}%
\providecommand \bibitemNoStop [0]{.\EOS\space}%
\providecommand \EOS [0]{\spacefactor3000\relax}%
\providecommand \BibitemShut  [1]{\csname bibitem#1\endcsname}%
\let\auto@bib@innerbib\@empty
%</preamble>
\bibitem [{\citenamefont {Loss}\ and\ \citenamefont {DiVincenzo}(1998)}]{Loss1998}%
  \BibitemOpen
  \bibfield  {author} {\bibinfo {author} {\bibfnamefont {D.}~\bibnamefont {Loss}}\ and\ \bibinfo {author} {\bibfnamefont {D.~P.}\ \bibnamefont {DiVincenzo}},\ }\bibfield  {title} {\bibinfo {title} {Quantum computation with quantum dots},\ }\href {https://doi.org/10.1103/physreva.57.120} {\bibfield  {journal} {\bibinfo  {journal} {Physical Review A}\ }\textbf {\bibinfo {volume} {57}},\ \bibinfo {pages} {120} (\bibinfo {year} {1998})}\BibitemShut {NoStop}%
\bibitem [{\citenamefont {Burkard}\ \emph {et~al.}(1999)\citenamefont {Burkard}, \citenamefont {Loss},\ and\ \citenamefont {DiVincenzo}}]{Burkard1999}%
  \BibitemOpen
  \bibfield  {author} {\bibinfo {author} {\bibfnamefont {G.}~\bibnamefont {Burkard}}, \bibinfo {author} {\bibfnamefont {D.}~\bibnamefont {Loss}},\ and\ \bibinfo {author} {\bibfnamefont {D.~P.}\ \bibnamefont {DiVincenzo}},\ }\bibfield  {title} {\bibinfo {title} {Coupled quantum dots as quantum gates},\ }\href {https://doi.org/10.1103/physrevb.59.2070} {\bibfield  {journal} {\bibinfo  {journal} {Physical Review B}\ }\textbf {\bibinfo {volume} {59}},\ \bibinfo {pages} {2070} (\bibinfo {year} {1999})}\BibitemShut {NoStop}%
\bibitem [{\citenamefont {Burkard}\ \emph {et~al.}(2023)\citenamefont {Burkard}, \citenamefont {Ladd}, \citenamefont {Nichol}, \citenamefont {Pan},\ and\ \citenamefont {Petta}}]{Burkard2023}%
  \BibitemOpen
  \bibfield  {author} {\bibinfo {author} {\bibfnamefont {G.}~\bibnamefont {Burkard}}, \bibinfo {author} {\bibfnamefont {T.~D.}\ \bibnamefont {Ladd}}, \bibinfo {author} {\bibfnamefont {J.~M.}\ \bibnamefont {Nichol}}, \bibinfo {author} {\bibfnamefont {A.}~\bibnamefont {Pan}},\ and\ \bibinfo {author} {\bibfnamefont {J.~R.}\ \bibnamefont {Petta}},\ }\bibfield  {title} {\bibinfo {title} {Semiconductor spin qubits},\ }\href {https://doi.org/10.1103/RevModPhys.95.025003} {\bibfield  {journal} {\bibinfo  {journal} {Rev. Mod. Phys.}\ }\textbf {\bibinfo {volume} {95}},\ \bibinfo {pages} {025003} (\bibinfo {year} {2023})},\ \Eprint {https://arxiv.org/abs/2112.08863} {arxiv:2112.08863} \BibitemShut {NoStop}%
\bibitem [{\citenamefont {Burkard}\ and\ \citenamefont {Imamo\u{g}lu}(2006)}]{Burkard2006}%
  \BibitemOpen
  \bibfield  {author} {\bibinfo {author} {\bibfnamefont {G.}~\bibnamefont {Burkard}}\ and\ \bibinfo {author} {\bibfnamefont {A.}~\bibnamefont {Imamo\u{g}lu}},\ }\bibfield  {title} {\bibinfo {title} {Ultra-long-distance interaction between spin qubits},\ }\href {https://doi.org/10.1103/PhysRevB.74.041307} {\bibfield  {journal} {\bibinfo  {journal} {Phys. Rev. B}\ }\textbf {\bibinfo {volume} {74}},\ \bibinfo {pages} {041307} (\bibinfo {year} {2006})},\ \Eprint {https://arxiv.org/abs/cond-mat/0603119} {cond-mat/0603119} \BibitemShut {NoStop}%
\bibitem [{\citenamefont {Trifunovic}\ \emph {et~al.}(2012)\citenamefont {Trifunovic}, \citenamefont {Dial}, \citenamefont {Trif}, \citenamefont {Wootton}, \citenamefont {Abebe}, \citenamefont {Yacoby},\ and\ \citenamefont {Loss}}]{Trifunovic2012}%
  \BibitemOpen
  \bibfield  {author} {\bibinfo {author} {\bibfnamefont {L.}~\bibnamefont {Trifunovic}}, \bibinfo {author} {\bibfnamefont {O.}~\bibnamefont {Dial}}, \bibinfo {author} {\bibfnamefont {M.}~\bibnamefont {Trif}}, \bibinfo {author} {\bibfnamefont {J.~R.}\ \bibnamefont {Wootton}}, \bibinfo {author} {\bibfnamefont {R.}~\bibnamefont {Abebe}}, \bibinfo {author} {\bibfnamefont {A.}~\bibnamefont {Yacoby}},\ and\ \bibinfo {author} {\bibfnamefont {D.}~\bibnamefont {Loss}},\ }\bibfield  {title} {\bibinfo {title} {Long-distance spin-spin coupling via floating gates},\ }\href {https://doi.org/10.1103/PhysRevX.2.011006} {\bibfield  {journal} {\bibinfo  {journal} {Phys. Rev. X}\ }\textbf {\bibinfo {volume} {2}},\ \bibinfo {pages} {011006} (\bibinfo {year} {2012})},\ \Eprint {https://arxiv.org/abs/1110.1342} {1110.1342} \BibitemShut {NoStop}%
\bibitem [{\citenamefont {Taylor}\ \emph {et~al.}(2005)\citenamefont {Taylor}, \citenamefont {Engel}, \citenamefont {D{\"u}r}, \citenamefont {Yacoby}, \citenamefont {Marcus}, \citenamefont {Zoller},\ and\ \citenamefont {Lukin}}]{Taylor2005}%
  \BibitemOpen
  \bibfield  {author} {\bibinfo {author} {\bibfnamefont {J.~M.}\ \bibnamefont {Taylor}}, \bibinfo {author} {\bibfnamefont {H.-A.}\ \bibnamefont {Engel}}, \bibinfo {author} {\bibfnamefont {W.}~\bibnamefont {D{\"u}r}}, \bibinfo {author} {\bibfnamefont {A.}~\bibnamefont {Yacoby}}, \bibinfo {author} {\bibfnamefont {C.~M.}\ \bibnamefont {Marcus}}, \bibinfo {author} {\bibfnamefont {P.}~\bibnamefont {Zoller}},\ and\ \bibinfo {author} {\bibfnamefont {M.~D.}\ \bibnamefont {Lukin}},\ }\bibfield  {title} {\bibinfo {title} {Fault-tolerant architecture for quantum computation using electrically controlled semiconductor spins},\ }\href {https://doi.org/10.1038/nphys174} {\bibfield  {journal} {\bibinfo  {journal} {Nature Phys.}\ }\textbf {\bibinfo {volume} {1}},\ \bibinfo {pages} {177} (\bibinfo {year} {2005})}\BibitemShut {NoStop}%
\bibitem [{\citenamefont {Barnes}\ \emph {et~al.}(2000)\citenamefont {Barnes}, \citenamefont {Shilton},\ and\ \citenamefont {Robinson}}]{Barnes2000}%
  \BibitemOpen
  \bibfield  {author} {\bibinfo {author} {\bibfnamefont {C.~H.~W.}\ \bibnamefont {Barnes}}, \bibinfo {author} {\bibfnamefont {J.~M.}\ \bibnamefont {Shilton}},\ and\ \bibinfo {author} {\bibfnamefont {A.~M.}\ \bibnamefont {Robinson}},\ }\bibfield  {title} {\bibinfo {title} {Quantum computation using electrons trapped by surface acoustic waves},\ }\href {https://doi.org/10.1103/physrevb.62.8410} {\bibfield  {journal} {\bibinfo  {journal} {Physical Review B}\ }\textbf {\bibinfo {volume} {62}},\ \bibinfo {pages} {8410} (\bibinfo {year} {2000})}\BibitemShut {NoStop}%
\bibitem [{\citenamefont {Pino}\ \emph {et~al.}(2021)\citenamefont {Pino}, \citenamefont {Dreiling}, \citenamefont {Figgatt}, \citenamefont {Gaebler}, \citenamefont {Moses}, \citenamefont {Allman}, \citenamefont {Baldwin}, \citenamefont {Foss-Feig}, \citenamefont {Hayes}, \citenamefont {Mayer}, \citenamefont {Ryan-Anderson},\ and\ \citenamefont {Neyenhuis}}]{Pino2021}%
  \BibitemOpen
  \bibfield  {author} {\bibinfo {author} {\bibfnamefont {J.~M.}\ \bibnamefont {Pino}}, \bibinfo {author} {\bibfnamefont {J.~M.}\ \bibnamefont {Dreiling}}, \bibinfo {author} {\bibfnamefont {C.}~\bibnamefont {Figgatt}}, \bibinfo {author} {\bibfnamefont {J.~P.}\ \bibnamefont {Gaebler}}, \bibinfo {author} {\bibfnamefont {S.~A.}\ \bibnamefont {Moses}}, \bibinfo {author} {\bibfnamefont {M.~S.}\ \bibnamefont {Allman}}, \bibinfo {author} {\bibfnamefont {C.~H.}\ \bibnamefont {Baldwin}}, \bibinfo {author} {\bibfnamefont {M.}~\bibnamefont {Foss-Feig}}, \bibinfo {author} {\bibfnamefont {D.}~\bibnamefont {Hayes}}, \bibinfo {author} {\bibfnamefont {K.}~\bibnamefont {Mayer}}, \bibinfo {author} {\bibfnamefont {C.}~\bibnamefont {Ryan-Anderson}},\ and\ \bibinfo {author} {\bibfnamefont {B.}~\bibnamefont {Neyenhuis}},\ }\bibfield  {title} {\bibinfo {title} {Demonstration of the trapped-ion quantum ccd computer architecture},\ }\href {https://doi.org/10.1038/s41586-021-03318-4} {\bibfield  {journal} {\bibinfo  {journal} {Nature}\
  }\textbf {\bibinfo {volume} {592}},\ \bibinfo {pages} {209–213} (\bibinfo {year} {2021})}\BibitemShut {NoStop}%
\bibitem [{\citenamefont {Sterk}\ \emph {et~al.}(2022)\citenamefont {Sterk}, \citenamefont {Coakley}, \citenamefont {Goldberg}, \citenamefont {Hietala}, \citenamefont {Lechtenberg}, \citenamefont {McGuinness}, \citenamefont {McMurtrey}, \citenamefont {Parazzoli}, \citenamefont {Van Der~Wall},\ and\ \citenamefont {Stick}}]{Sterk2022}%
  \BibitemOpen
  \bibfield  {author} {\bibinfo {author} {\bibfnamefont {J.~D.}\ \bibnamefont {Sterk}}, \bibinfo {author} {\bibfnamefont {H.}~\bibnamefont {Coakley}}, \bibinfo {author} {\bibfnamefont {J.}~\bibnamefont {Goldberg}}, \bibinfo {author} {\bibfnamefont {V.}~\bibnamefont {Hietala}}, \bibinfo {author} {\bibfnamefont {J.}~\bibnamefont {Lechtenberg}}, \bibinfo {author} {\bibfnamefont {H.}~\bibnamefont {McGuinness}}, \bibinfo {author} {\bibfnamefont {D.}~\bibnamefont {McMurtrey}}, \bibinfo {author} {\bibfnamefont {L.~P.}\ \bibnamefont {Parazzoli}}, \bibinfo {author} {\bibfnamefont {J.}~\bibnamefont {Van Der~Wall}},\ and\ \bibinfo {author} {\bibfnamefont {D.}~\bibnamefont {Stick}},\ }\bibfield  {title} {\bibinfo {title} {Closed-loop optimization of fast trapped-ion shuttling with sub-quanta excitation},\ }\bibfield  {journal} {\bibinfo  {journal} {npj Quantum Information}\ }\textbf {\bibinfo {volume} {8}},\ \href {https://doi.org/10.1038/s41534-022-00579-3} {10.1038/s41534-022-00579-3} (\bibinfo {year}
  {2022})\BibitemShut {NoStop}%
\bibitem [{\citenamefont {Bluvstein}\ \emph {et~al.}(2022)\citenamefont {Bluvstein}, \citenamefont {Levine}, \citenamefont {Semeghini}, \citenamefont {Wang}, \citenamefont {Ebadi}, \citenamefont {Kalinowski}, \citenamefont {Keesling}, \citenamefont {Maskara}, \citenamefont {Pichler}, \citenamefont {Greiner}, \citenamefont {Vuletić},\ and\ \citenamefont {Lukin}}]{Bluvstein2022}%
  \BibitemOpen
  \bibfield  {author} {\bibinfo {author} {\bibfnamefont {D.}~\bibnamefont {Bluvstein}}, \bibinfo {author} {\bibfnamefont {H.}~\bibnamefont {Levine}}, \bibinfo {author} {\bibfnamefont {G.}~\bibnamefont {Semeghini}}, \bibinfo {author} {\bibfnamefont {T.~T.}\ \bibnamefont {Wang}}, \bibinfo {author} {\bibfnamefont {S.}~\bibnamefont {Ebadi}}, \bibinfo {author} {\bibfnamefont {M.}~\bibnamefont {Kalinowski}}, \bibinfo {author} {\bibfnamefont {A.}~\bibnamefont {Keesling}}, \bibinfo {author} {\bibfnamefont {N.}~\bibnamefont {Maskara}}, \bibinfo {author} {\bibfnamefont {H.}~\bibnamefont {Pichler}}, \bibinfo {author} {\bibfnamefont {M.}~\bibnamefont {Greiner}}, \bibinfo {author} {\bibfnamefont {V.}~\bibnamefont {Vuletić}},\ and\ \bibinfo {author} {\bibfnamefont {M.~D.}\ \bibnamefont {Lukin}},\ }\bibfield  {title} {\bibinfo {title} {A quantum processor based on coherent transport of entangled atom arrays},\ }\href {https://doi.org/10.1038/s41586-022-04592-6} {\bibfield  {journal} {\bibinfo  {journal} {Nature}\ }\textbf
  {\bibinfo {volume} {604}},\ \bibinfo {pages} {451–456} (\bibinfo {year} {2022})}\BibitemShut {NoStop}%
\bibitem [{\citenamefont {Bluvstein}\ \emph {et~al.}(2023)\citenamefont {Bluvstein}, \citenamefont {Evered}, \citenamefont {Geim}, \citenamefont {Li}, \citenamefont {Zhou}, \citenamefont {Manovitz}, \citenamefont {Ebadi}, \citenamefont {Cain}, \citenamefont {Kalinowski}, \citenamefont {Hangleiter}, \citenamefont {Bonilla~Ataides}, \citenamefont {Maskara}, \citenamefont {Cong}, \citenamefont {Gao}, \citenamefont {Sales~Rodriguez}, \citenamefont {Karolyshyn}, \citenamefont {Semeghini}, \citenamefont {Gullans}, \citenamefont {Greiner}, \citenamefont {Vuletić},\ and\ \citenamefont {Lukin}}]{Bluvstein2023}%
  \BibitemOpen
  \bibfield  {author} {\bibinfo {author} {\bibfnamefont {D.}~\bibnamefont {Bluvstein}}, \bibinfo {author} {\bibfnamefont {S.~J.}\ \bibnamefont {Evered}}, \bibinfo {author} {\bibfnamefont {A.~A.}\ \bibnamefont {Geim}}, \bibinfo {author} {\bibfnamefont {S.~H.}\ \bibnamefont {Li}}, \bibinfo {author} {\bibfnamefont {H.}~\bibnamefont {Zhou}}, \bibinfo {author} {\bibfnamefont {T.}~\bibnamefont {Manovitz}}, \bibinfo {author} {\bibfnamefont {S.}~\bibnamefont {Ebadi}}, \bibinfo {author} {\bibfnamefont {M.}~\bibnamefont {Cain}}, \bibinfo {author} {\bibfnamefont {M.}~\bibnamefont {Kalinowski}}, \bibinfo {author} {\bibfnamefont {D.}~\bibnamefont {Hangleiter}}, \bibinfo {author} {\bibfnamefont {J.~P.}\ \bibnamefont {Bonilla~Ataides}}, \bibinfo {author} {\bibfnamefont {N.}~\bibnamefont {Maskara}}, \bibinfo {author} {\bibfnamefont {I.}~\bibnamefont {Cong}}, \bibinfo {author} {\bibfnamefont {X.}~\bibnamefont {Gao}}, \bibinfo {author} {\bibfnamefont {P.}~\bibnamefont {Sales~Rodriguez}}, \bibinfo {author} {\bibfnamefont
  {T.}~\bibnamefont {Karolyshyn}}, \bibinfo {author} {\bibfnamefont {G.}~\bibnamefont {Semeghini}}, \bibinfo {author} {\bibfnamefont {M.~J.}\ \bibnamefont {Gullans}}, \bibinfo {author} {\bibfnamefont {M.}~\bibnamefont {Greiner}}, \bibinfo {author} {\bibfnamefont {V.}~\bibnamefont {Vuletić}},\ and\ \bibinfo {author} {\bibfnamefont {M.~D.}\ \bibnamefont {Lukin}},\ }\bibfield  {title} {\bibinfo {title} {Logical quantum processor based on reconfigurable atom arrays},\ }\href {https://doi.org/10.1038/s41586-023-06927-3} {\bibfield  {journal} {\bibinfo  {journal} {Nature}\ }\textbf {\bibinfo {volume} {626}},\ \bibinfo {pages} {58–65} (\bibinfo {year} {2023})}\BibitemShut {NoStop}%
\bibitem [{\citenamefont {Mills}\ \emph {et~al.}(2019)\citenamefont {Mills}, \citenamefont {Zajac}, \citenamefont {Gullans}, \citenamefont {Schupp}, \citenamefont {Hazard},\ and\ \citenamefont {Petta}}]{Mills2019}%
  \BibitemOpen
  \bibfield  {author} {\bibinfo {author} {\bibfnamefont {A.~R.}\ \bibnamefont {Mills}}, \bibinfo {author} {\bibfnamefont {D.~M.}\ \bibnamefont {Zajac}}, \bibinfo {author} {\bibfnamefont {M.~J.}\ \bibnamefont {Gullans}}, \bibinfo {author} {\bibfnamefont {F.~J.}\ \bibnamefont {Schupp}}, \bibinfo {author} {\bibfnamefont {T.~M.}\ \bibnamefont {Hazard}},\ and\ \bibinfo {author} {\bibfnamefont {J.~R.}\ \bibnamefont {Petta}},\ }\bibfield  {title} {\bibinfo {title} {Shuttling a single charge across a one-dimensional array of silicon quantum dots},\ }\bibfield  {journal} {\bibinfo  {journal} {Nature Communications}\ }\textbf {\bibinfo {volume} {10}},\ \href {https://doi.org/10.1038/s41467-019-08970-z} {10.1038/s41467-019-08970-z} (\bibinfo {year} {2019})\BibitemShut {NoStop}%
\bibitem [{\citenamefont {Zwerver}\ \emph {et~al.}(2023)\citenamefont {Zwerver}, \citenamefont {Amitonov}, \citenamefont {de~Snoo}, \citenamefont {Mądzik}, \citenamefont {Rimbach-Russ}, \citenamefont {Sammak}, \citenamefont {Scappucci},\ and\ \citenamefont {Vandersypen}}]{Zwerver2023}%
  \BibitemOpen
  \bibfield  {author} {\bibinfo {author} {\bibfnamefont {A.~M.~J.}\ \bibnamefont {Zwerver}}, \bibinfo {author} {\bibfnamefont {S.~V.}\ \bibnamefont {Amitonov}}, \bibinfo {author} {\bibfnamefont {S.~L.}\ \bibnamefont {de~Snoo}}, \bibinfo {author} {\bibfnamefont {M.~T.}\ \bibnamefont {Mądzik}}, \bibinfo {author} {\bibfnamefont {M.}~\bibnamefont {Rimbach-Russ}}, \bibinfo {author} {\bibfnamefont {A.}~\bibnamefont {Sammak}}, \bibinfo {author} {\bibfnamefont {G.}~\bibnamefont {Scappucci}},\ and\ \bibinfo {author} {\bibfnamefont {L.~M.~K.}\ \bibnamefont {Vandersypen}},\ }\bibfield  {title} {\bibinfo {title} {Shuttling an electron spin through a silicon quantum dot array},\ }\bibfield  {journal} {\bibinfo  {journal} {PRX Quantum}\ }\textbf {\bibinfo {volume} {4}},\ \href {https://doi.org/10.1103/prxquantum.4.030303} {10.1103/prxquantum.4.030303} (\bibinfo {year} {2023})\BibitemShut {NoStop}%
\bibitem [{\citenamefont {Matsumoto}\ \emph {et~al.}(2025)\citenamefont {Matsumoto}, \citenamefont {De~Smet}, \citenamefont {Tryputen}, \citenamefont {de~Snoo}, \citenamefont {Amitonov}, \citenamefont {Sammak}, \citenamefont {Rimbach-Russ}, \citenamefont {Scappucci},\ and\ \citenamefont {Vandersypen}}]{Matsumoto2025}%
  \BibitemOpen
  \bibfield  {author} {\bibinfo {author} {\bibfnamefont {Y.}~\bibnamefont {Matsumoto}}, \bibinfo {author} {\bibfnamefont {M.}~\bibnamefont {De~Smet}}, \bibinfo {author} {\bibfnamefont {L.}~\bibnamefont {Tryputen}}, \bibinfo {author} {\bibfnamefont {S.~L.}\ \bibnamefont {de~Snoo}}, \bibinfo {author} {\bibfnamefont {S.~V.}\ \bibnamefont {Amitonov}}, \bibinfo {author} {\bibfnamefont {A.}~\bibnamefont {Sammak}}, \bibinfo {author} {\bibfnamefont {M.}~\bibnamefont {Rimbach-Russ}}, \bibinfo {author} {\bibfnamefont {G.}~\bibnamefont {Scappucci}},\ and\ \bibinfo {author} {\bibfnamefont {L.~M.~K.}\ \bibnamefont {Vandersypen}},\ }\href {https://doi.org/10.48550/ARXIV.2503.15434} {\bibinfo {title} {Two-qubit logic and teleportation with mobile spin qubits in silicon}} (\bibinfo {year} {2025})\BibitemShut {NoStop}%
\bibitem [{\citenamefont {McNeil}\ \emph {et~al.}(2011)\citenamefont {McNeil}, \citenamefont {Kataoka}, \citenamefont {Ford}, \citenamefont {Barnes}, \citenamefont {Anderson}, \citenamefont {Jones}, \citenamefont {Farrer},\ and\ \citenamefont {Ritchie}}]{McNeil2011}%
  \BibitemOpen
  \bibfield  {author} {\bibinfo {author} {\bibfnamefont {R.~P.~G.}\ \bibnamefont {McNeil}}, \bibinfo {author} {\bibfnamefont {M.}~\bibnamefont {Kataoka}}, \bibinfo {author} {\bibfnamefont {C.~J.~B.}\ \bibnamefont {Ford}}, \bibinfo {author} {\bibfnamefont {C.~H.~W.}\ \bibnamefont {Barnes}}, \bibinfo {author} {\bibfnamefont {D.}~\bibnamefont {Anderson}}, \bibinfo {author} {\bibfnamefont {G.~A.~C.}\ \bibnamefont {Jones}}, \bibinfo {author} {\bibfnamefont {I.}~\bibnamefont {Farrer}},\ and\ \bibinfo {author} {\bibfnamefont {D.~A.}\ \bibnamefont {Ritchie}},\ }\bibfield  {title} {\bibinfo {title} {On-demand single-electron transfer between distant quantum dots},\ }\href {https://doi.org/10.1038/nature10444} {\bibfield  {journal} {\bibinfo  {journal} {Nature}\ }\textbf {\bibinfo {volume} {477}},\ \bibinfo {pages} {439–442} (\bibinfo {year} {2011})}\BibitemShut {NoStop}%
\bibitem [{\citenamefont {Ford}(2017)}]{Ford2017}%
  \BibitemOpen
  \bibfield  {author} {\bibinfo {author} {\bibfnamefont {C.~J.~B.}\ \bibnamefont {Ford}},\ }\bibfield  {title} {\bibinfo {title} {Transporting and manipulating single electrons in surface-acoustic-wave minima},\ }\href {https://doi.org/10.1002/pssb.201600658} {\bibfield  {journal} {\bibinfo  {journal} {physica status solidi (b)}\ }\textbf {\bibinfo {volume} {254}},\ \bibinfo {pages} {1600658} (\bibinfo {year} {2017})}\BibitemShut {NoStop}%
\bibitem [{\citenamefont {Vandersypen}\ and\ \citenamefont {Eriksson}(2019)}]{Vandersypen2019}%
  \BibitemOpen
  \bibfield  {author} {\bibinfo {author} {\bibfnamefont {L.~M.~K.}\ \bibnamefont {Vandersypen}}\ and\ \bibinfo {author} {\bibfnamefont {M.~A.}\ \bibnamefont {Eriksson}},\ }\bibfield  {title} {\bibinfo {title} {Quantum computing with semiconductor spins},\ }\href {https://doi.org/10.1063/pt.3.4270} {\bibfield  {journal} {\bibinfo  {journal} {Physics Today}\ }\textbf {\bibinfo {volume} {72}},\ \bibinfo {pages} {38–45} (\bibinfo {year} {2019})}\BibitemShut {NoStop}%
\bibitem [{\citenamefont {Zwerver}\ \emph {et~al.}(2022)\citenamefont {Zwerver}, \citenamefont {Kr\"{a}henmann}, \citenamefont {Watson}, \citenamefont {Lampert}, \citenamefont {George}, \citenamefont {Pillarisetty}, \citenamefont {Bojarski}, \citenamefont {Amin}, \citenamefont {Amitonov}, \citenamefont {Boter}, \citenamefont {Caudillo}, \citenamefont {Correas-Serrano}, \citenamefont {Dehollain}, \citenamefont {Droulers}, \citenamefont {Henry}, \citenamefont {Kotlyar}, \citenamefont {Lodari}, \citenamefont {L\"{u}thi}, \citenamefont {Michalak}, \citenamefont {Mueller}, \citenamefont {Neyens}, \citenamefont {Roberts}, \citenamefont {Samkharadze}, \citenamefont {Zheng}, \citenamefont {Zietz}, \citenamefont {Scappucci}, \citenamefont {Veldhorst}, \citenamefont {Vandersypen},\ and\ \citenamefont {Clarke}}]{Zwerver2022}%
  \BibitemOpen
  \bibfield  {author} {\bibinfo {author} {\bibfnamefont {A.~M.~J.}\ \bibnamefont {Zwerver}}, \bibinfo {author} {\bibfnamefont {T.}~\bibnamefont {Kr\"{a}henmann}}, \bibinfo {author} {\bibfnamefont {T.~F.}\ \bibnamefont {Watson}}, \bibinfo {author} {\bibfnamefont {L.}~\bibnamefont {Lampert}}, \bibinfo {author} {\bibfnamefont {H.~C.}\ \bibnamefont {George}}, \bibinfo {author} {\bibfnamefont {R.}~\bibnamefont {Pillarisetty}}, \bibinfo {author} {\bibfnamefont {S.~A.}\ \bibnamefont {Bojarski}}, \bibinfo {author} {\bibfnamefont {P.}~\bibnamefont {Amin}}, \bibinfo {author} {\bibfnamefont {S.~V.}\ \bibnamefont {Amitonov}}, \bibinfo {author} {\bibfnamefont {J.~M.}\ \bibnamefont {Boter}}, \bibinfo {author} {\bibfnamefont {R.}~\bibnamefont {Caudillo}}, \bibinfo {author} {\bibfnamefont {D.}~\bibnamefont {Correas-Serrano}}, \bibinfo {author} {\bibfnamefont {J.~P.}\ \bibnamefont {Dehollain}}, \bibinfo {author} {\bibfnamefont {G.}~\bibnamefont {Droulers}}, \bibinfo {author} {\bibfnamefont {E.~M.}\ \bibnamefont {Henry}},
  \bibinfo {author} {\bibfnamefont {R.}~\bibnamefont {Kotlyar}}, \bibinfo {author} {\bibfnamefont {M.}~\bibnamefont {Lodari}}, \bibinfo {author} {\bibfnamefont {F.}~\bibnamefont {L\"{u}thi}}, \bibinfo {author} {\bibfnamefont {D.~J.}\ \bibnamefont {Michalak}}, \bibinfo {author} {\bibfnamefont {B.~K.}\ \bibnamefont {Mueller}}, \bibinfo {author} {\bibfnamefont {S.}~\bibnamefont {Neyens}}, \bibinfo {author} {\bibfnamefont {J.}~\bibnamefont {Roberts}}, \bibinfo {author} {\bibfnamefont {N.}~\bibnamefont {Samkharadze}}, \bibinfo {author} {\bibfnamefont {G.}~\bibnamefont {Zheng}}, \bibinfo {author} {\bibfnamefont {O.~K.}\ \bibnamefont {Zietz}}, \bibinfo {author} {\bibfnamefont {G.}~\bibnamefont {Scappucci}}, \bibinfo {author} {\bibfnamefont {M.}~\bibnamefont {Veldhorst}}, \bibinfo {author} {\bibfnamefont {L.~M.~K.}\ \bibnamefont {Vandersypen}},\ and\ \bibinfo {author} {\bibfnamefont {J.~S.}\ \bibnamefont {Clarke}},\ }\bibfield  {title} {\bibinfo {title} {Qubits made by advanced semiconductor manufacturing},\ }\href
  {https://doi.org/10.1038/s41928-022-00727-9} {\bibfield  {journal} {\bibinfo  {journal} {Nature Electronics}\ }\textbf {\bibinfo {volume} {5}},\ \bibinfo {pages} {184–190} (\bibinfo {year} {2022})}\BibitemShut {NoStop}%
\bibitem [{\citenamefont {Wang}\ \emph {et~al.}(2024)\citenamefont {Wang}, \citenamefont {Edlbauer}, \citenamefont {Jadot}, \citenamefont {Meunier}, \citenamefont {Takada}, \citenamefont {B\"{a}uerle},\ and\ \citenamefont {Sellier}}]{Wang2024}%
  \BibitemOpen
  \bibfield  {author} {\bibinfo {author} {\bibfnamefont {J.}~\bibnamefont {Wang}}, \bibinfo {author} {\bibfnamefont {H.}~\bibnamefont {Edlbauer}}, \bibinfo {author} {\bibfnamefont {B.}~\bibnamefont {Jadot}}, \bibinfo {author} {\bibfnamefont {T.}~\bibnamefont {Meunier}}, \bibinfo {author} {\bibfnamefont {S.}~\bibnamefont {Takada}}, \bibinfo {author} {\bibfnamefont {C.}~\bibnamefont {B\"{a}uerle}},\ and\ \bibinfo {author} {\bibfnamefont {H.}~\bibnamefont {Sellier}},\ }\bibfield  {title} {\bibinfo {title} {Electron qubits surfing on acoustic waves: review of recent progress},\ }\href {https://doi.org/10.1088/1361-6463/ad6c5a} {\bibfield  {journal} {\bibinfo  {journal} {Journal of Physics D: Applied Physics}\ }\textbf {\bibinfo {volume} {58}},\ \bibinfo {pages} {023002} (\bibinfo {year} {2024})}\BibitemShut {NoStop}%
\bibitem [{\citenamefont {van~der Wiel}\ \emph {et~al.}(2002)\citenamefont {van~der Wiel}, \citenamefont {De~Franceschi}, \citenamefont {Elzerman}, \citenamefont {Fujisawa}, \citenamefont {Tarucha},\ and\ \citenamefont {Kouwenhoven}}]{vanderWiel2002}%
  \BibitemOpen
  \bibfield  {author} {\bibinfo {author} {\bibfnamefont {W.~G.}\ \bibnamefont {van~der Wiel}}, \bibinfo {author} {\bibfnamefont {S.}~\bibnamefont {De~Franceschi}}, \bibinfo {author} {\bibfnamefont {J.~M.}\ \bibnamefont {Elzerman}}, \bibinfo {author} {\bibfnamefont {T.}~\bibnamefont {Fujisawa}}, \bibinfo {author} {\bibfnamefont {S.}~\bibnamefont {Tarucha}},\ and\ \bibinfo {author} {\bibfnamefont {L.~P.}\ \bibnamefont {Kouwenhoven}},\ }\bibfield  {title} {\bibinfo {title} {Electron transport through double quantum dots},\ }\href {https://doi.org/10.1103/revmodphys.75.1} {\bibfield  {journal} {\bibinfo  {journal} {Reviews of Modern Physics}\ }\textbf {\bibinfo {volume} {75}},\ \bibinfo {pages} {1–22} (\bibinfo {year} {2002})}\BibitemShut {NoStop}%
\bibitem [{\citenamefont {Langrock}\ \emph {et~al.}(2023)\citenamefont {Langrock}, \citenamefont {Krzywda}, \citenamefont {Focke}, \citenamefont {Seidler}, \citenamefont {Schreiber},\ and\ \citenamefont {Cywi{\'n}ski}}]{Langrock2023}%
  \BibitemOpen
  \bibfield  {author} {\bibinfo {author} {\bibfnamefont {V.}~\bibnamefont {Langrock}}, \bibinfo {author} {\bibfnamefont {J.~A.}\ \bibnamefont {Krzywda}}, \bibinfo {author} {\bibfnamefont {N.}~\bibnamefont {Focke}}, \bibinfo {author} {\bibfnamefont {I.}~\bibnamefont {Seidler}}, \bibinfo {author} {\bibfnamefont {L.~R.}\ \bibnamefont {Schreiber}},\ and\ \bibinfo {author} {\bibfnamefont {{\L}.}~\bibnamefont {Cywi{\'n}ski}},\ }\bibfield  {title} {{\selectlanguage {english}\bibinfo {title} {Blueprint of a scalable spin qubit shuttle device for coherent mid-range qubit transfer in disordered {Si/SiGe/SiO2}}},\ }\href@noop {} {\bibfield  {journal} {\bibinfo  {journal} {PRX quantum}\ }\textbf {\bibinfo {volume} {4}} (\bibinfo {year} {2023})}\BibitemShut {NoStop}%
\bibitem [{\citenamefont {Buonacorsi}\ \emph {et~al.}(2020)\citenamefont {Buonacorsi}, \citenamefont {Shaw},\ and\ \citenamefont {Baugh}}]{Buonacorsi2020}%
  \BibitemOpen
  \bibfield  {author} {\bibinfo {author} {\bibfnamefont {B.}~\bibnamefont {Buonacorsi}}, \bibinfo {author} {\bibfnamefont {B.}~\bibnamefont {Shaw}},\ and\ \bibinfo {author} {\bibfnamefont {J.}~\bibnamefont {Baugh}},\ }\bibfield  {title} {\bibinfo {title} {Simulated coherent electron shuttling in silicon quantum dots},\ }\bibfield  {journal} {\bibinfo  {journal} {Physical Review B}\ }\textbf {\bibinfo {volume} {102}},\ \href {https://doi.org/10.1103/physrevb.102.125406} {10.1103/physrevb.102.125406} (\bibinfo {year} {2020})\BibitemShut {NoStop}%
\bibitem [{\citenamefont {Machnes}\ \emph {et~al.}(2011)\citenamefont {Machnes}, \citenamefont {Sander}, \citenamefont {Glaser}, \citenamefont {de~Fouquières}, \citenamefont {Gruslys}, \citenamefont {Schirmer},\ and\ \citenamefont {Schulte-Herbr\"{u}ggen}}]{Machnes2011}%
  \BibitemOpen
  \bibfield  {author} {\bibinfo {author} {\bibfnamefont {S.}~\bibnamefont {Machnes}}, \bibinfo {author} {\bibfnamefont {U.}~\bibnamefont {Sander}}, \bibinfo {author} {\bibfnamefont {S.~J.}\ \bibnamefont {Glaser}}, \bibinfo {author} {\bibfnamefont {P.}~\bibnamefont {de~Fouquières}}, \bibinfo {author} {\bibfnamefont {A.}~\bibnamefont {Gruslys}}, \bibinfo {author} {\bibfnamefont {S.}~\bibnamefont {Schirmer}},\ and\ \bibinfo {author} {\bibfnamefont {T.}~\bibnamefont {Schulte-Herbr\"{u}ggen}},\ }\bibfield  {title} {\bibinfo {title} {Comparing, optimizing, and benchmarking quantum-control algorithms in a unifying programming framework},\ }\bibfield  {journal} {\bibinfo  {journal} {Physical Review A}\ }\textbf {\bibinfo {volume} {84}},\ \href {https://doi.org/10.1103/physreva.84.022305} {10.1103/physreva.84.022305} (\bibinfo {year} {2011})\BibitemShut {NoStop}%
\bibitem [{\citenamefont {Mortensen}\ \emph {et~al.}(2018)\citenamefont {Mortensen}, \citenamefont {S{\o}rensen}, \citenamefont {M{\o}lmer},\ and\ \citenamefont {Sherson}}]{Mortensen2018}%
  \BibitemOpen
  \bibfield  {author} {\bibinfo {author} {\bibfnamefont {H.~L.}\ \bibnamefont {Mortensen}}, \bibinfo {author} {\bibfnamefont {J.~J. W.~H.}\ \bibnamefont {S{\o}rensen}}, \bibinfo {author} {\bibfnamefont {K.}~\bibnamefont {M{\o}lmer}},\ and\ \bibinfo {author} {\bibfnamefont {J.~F.}\ \bibnamefont {Sherson}},\ }\bibfield  {title} {\bibinfo {title} {Fast state transfer in a {$\Lambda$-system}: a shortcut-to-adiabaticity approach to robust and resource optimized control},\ }\href@noop {} {\bibfield  {journal} {\bibinfo  {journal} {New J. Phys.}\ }\textbf {\bibinfo {volume} {20}},\ \bibinfo {pages} {025009} (\bibinfo {year} {2018})}\BibitemShut {NoStop}%
\bibitem [{\citenamefont {Guéry-Odelin}\ \emph {et~al.}(2019)\citenamefont {Guéry-Odelin}, \citenamefont {Ruschhaupt}, \citenamefont {Kiely}, \citenamefont {Torrontegui}, \citenamefont {Mart\'{i}nez-Garaot},\ and\ \citenamefont {Muga}}]{GuryOdelin2019}%
  \BibitemOpen
  \bibfield  {author} {\bibinfo {author} {\bibfnamefont {D.}~\bibnamefont {Guéry-Odelin}}, \bibinfo {author} {\bibfnamefont {A.}~\bibnamefont {Ruschhaupt}}, \bibinfo {author} {\bibfnamefont {A.}~\bibnamefont {Kiely}}, \bibinfo {author} {\bibfnamefont {E.}~\bibnamefont {Torrontegui}}, \bibinfo {author} {\bibfnamefont {S.}~\bibnamefont {Mart\'{i}nez-Garaot}},\ and\ \bibinfo {author} {\bibfnamefont {J.}~\bibnamefont {Muga}},\ }\bibfield  {title} {\bibinfo {title} {Shortcuts to adiabaticity: Concepts, methods, and applications},\ }\bibfield  {journal} {\bibinfo  {journal} {Reviews of Modern Physics}\ }\textbf {\bibinfo {volume} {91}},\ \href {https://doi.org/10.1103/revmodphys.91.045001} {10.1103/revmodphys.91.045001} (\bibinfo {year} {2019})\BibitemShut {NoStop}%
\bibitem [{\citenamefont {Wang}\ \emph {et~al.}(2022)\citenamefont {Wang}, \citenamefont {Ota}, \citenamefont {Edlbauer}, \citenamefont {Jadot}, \citenamefont {Mortemousque}, \citenamefont {Richard}, \citenamefont {Okazaki}, \citenamefont {Nakamura}, \citenamefont {Ludwig}, \citenamefont {Wieck}, \citenamefont {Urdampilleta}, \citenamefont {Meunier}, \citenamefont {Kodera}, \citenamefont {Kaneko}, \citenamefont {Takada},\ and\ \citenamefont {B\"{a}uerle}}]{Wang2022}%
  \BibitemOpen
  \bibfield  {author} {\bibinfo {author} {\bibfnamefont {J.}~\bibnamefont {Wang}}, \bibinfo {author} {\bibfnamefont {S.}~\bibnamefont {Ota}}, \bibinfo {author} {\bibfnamefont {H.}~\bibnamefont {Edlbauer}}, \bibinfo {author} {\bibfnamefont {B.}~\bibnamefont {Jadot}}, \bibinfo {author} {\bibfnamefont {P.-A.}\ \bibnamefont {Mortemousque}}, \bibinfo {author} {\bibfnamefont {A.}~\bibnamefont {Richard}}, \bibinfo {author} {\bibfnamefont {Y.}~\bibnamefont {Okazaki}}, \bibinfo {author} {\bibfnamefont {S.}~\bibnamefont {Nakamura}}, \bibinfo {author} {\bibfnamefont {A.}~\bibnamefont {Ludwig}}, \bibinfo {author} {\bibfnamefont {A.~D.}\ \bibnamefont {Wieck}}, \bibinfo {author} {\bibfnamefont {M.}~\bibnamefont {Urdampilleta}}, \bibinfo {author} {\bibfnamefont {T.}~\bibnamefont {Meunier}}, \bibinfo {author} {\bibfnamefont {T.}~\bibnamefont {Kodera}}, \bibinfo {author} {\bibfnamefont {N.-H.}\ \bibnamefont {Kaneko}}, \bibinfo {author} {\bibfnamefont {S.}~\bibnamefont {Takada}},\ and\ \bibinfo {author} {\bibfnamefont
  {C.}~\bibnamefont {B\"{a}uerle}},\ }\bibfield  {title} {\bibinfo {title} {Generation of a single-cycle acoustic pulse: A scalable solution for transport in single-electron circuits},\ }\bibfield  {journal} {\bibinfo  {journal} {Physical Review X}\ }\textbf {\bibinfo {volume} {12}},\ \href {https://doi.org/10.1103/physrevx.12.031035} {10.1103/physrevx.12.031035} (\bibinfo {year} {2022})\BibitemShut {NoStop}%
\bibitem [{\citenamefont {Furuta}\ \emph {et~al.}(2004)\citenamefont {Furuta}, \citenamefont {Barnes},\ and\ \citenamefont {Doran}}]{Doran2004}%
  \BibitemOpen
  \bibfield  {author} {\bibinfo {author} {\bibfnamefont {S.}~\bibnamefont {Furuta}}, \bibinfo {author} {\bibfnamefont {C.~H.~W.}\ \bibnamefont {Barnes}},\ and\ \bibinfo {author} {\bibfnamefont {C.~J.~L.}\ \bibnamefont {Doran}},\ }\bibfield  {title} {\bibinfo {title} {Single-qubit gates and measurements in the surface acoustic wave quantum computer},\ }\href {https://doi.org/10.1103/PhysRevB.70.205320} {\bibfield  {journal} {\bibinfo  {journal} {Phys. Rev. B}\ }\textbf {\bibinfo {volume} {70}},\ \bibinfo {pages} {205320} (\bibinfo {year} {2004})}\BibitemShut {NoStop}%
\bibitem [{\citenamefont {Bäuerle}\ \emph {et~al.}(2018)\citenamefont {Bäuerle}, \citenamefont {Glattli}, \citenamefont {Meunier}, \citenamefont {Portier}, \citenamefont {Roche}, \citenamefont {Roulleau}, \citenamefont {Takada},\ and\ \citenamefont {Waintal}}]{Bauerle2018}%
  \BibitemOpen
  \bibfield  {author} {\bibinfo {author} {\bibfnamefont {C.}~\bibnamefont {Bäuerle}}, \bibinfo {author} {\bibfnamefont {D.~C.}\ \bibnamefont {Glattli}}, \bibinfo {author} {\bibfnamefont {T.}~\bibnamefont {Meunier}}, \bibinfo {author} {\bibfnamefont {F.}~\bibnamefont {Portier}}, \bibinfo {author} {\bibfnamefont {P.}~\bibnamefont {Roche}}, \bibinfo {author} {\bibfnamefont {P.}~\bibnamefont {Roulleau}}, \bibinfo {author} {\bibfnamefont {S.}~\bibnamefont {Takada}},\ and\ \bibinfo {author} {\bibfnamefont {X.}~\bibnamefont {Waintal}},\ }\bibfield  {title} {\bibinfo {title} {Coherent control of single electrons: a review of current progress},\ }\href {https://doi.org/10.1088/1361-6633/aaa98a} {\bibfield  {journal} {\bibinfo  {journal} {Reports on Progress in Physics}\ }\textbf {\bibinfo {volume} {81}},\ \bibinfo {pages} {056503} (\bibinfo {year} {2018})}\BibitemShut {NoStop}%
\bibitem [{\citenamefont {Lepage}\ \emph {et~al.}(2020)\citenamefont {Lepage}, \citenamefont {Lasek}, \citenamefont {Arvidsson-Shukur},\ and\ \citenamefont {Barnes}}]{Lepage2020}%
  \BibitemOpen
  \bibfield  {author} {\bibinfo {author} {\bibfnamefont {H.~V.}\ \bibnamefont {Lepage}}, \bibinfo {author} {\bibfnamefont {A.~A.}\ \bibnamefont {Lasek}}, \bibinfo {author} {\bibfnamefont {D.~R.~M.}\ \bibnamefont {Arvidsson-Shukur}},\ and\ \bibinfo {author} {\bibfnamefont {C.~H.~W.}\ \bibnamefont {Barnes}},\ }\bibfield  {title} {\bibinfo {title} {Entanglement generation via power-of-swap operations between dynamic electron-spin qubits},\ }\bibfield  {journal} {\bibinfo  {journal} {Physical Review A}\ }\textbf {\bibinfo {volume} {101}},\ \href {https://doi.org/10.1103/physreva.101.022329} {10.1103/physreva.101.022329} (\bibinfo {year} {2020})\BibitemShut {NoStop}%
\bibitem [{\citenamefont {Benito}\ \emph {et~al.}(2016)\citenamefont {Benito}, \citenamefont {Schuetz}, \citenamefont {Cirac}, \citenamefont {Platero},\ and\ \citenamefont {Giedke}}]{Benito2016}%
  \BibitemOpen
  \bibfield  {author} {\bibinfo {author} {\bibfnamefont {M.}~\bibnamefont {Benito}}, \bibinfo {author} {\bibfnamefont {M.~J.~A.}\ \bibnamefont {Schuetz}}, \bibinfo {author} {\bibfnamefont {J.~I.}\ \bibnamefont {Cirac}}, \bibinfo {author} {\bibfnamefont {G.}~\bibnamefont {Platero}},\ and\ \bibinfo {author} {\bibfnamefont {G.}~\bibnamefont {Giedke}},\ }\bibfield  {title} {\bibinfo {title} {Dissipative long-range entanglement generation between electronic spins},\ }\href {https://doi.org/10.1103/PhysRevB.94.115404} {\bibfield  {journal} {\bibinfo  {journal} {Phys. Rev. B}\ }\textbf {\bibinfo {volume} {94}},\ \bibinfo {pages} {115404} (\bibinfo {year} {2016})}\BibitemShut {NoStop}%
\bibitem [{\citenamefont {Jadot}\ \emph {et~al.}(2021)\citenamefont {Jadot}, \citenamefont {Mortemousque}, \citenamefont {Chanrion}, \citenamefont {Thiney}, \citenamefont {Ludwig}, \citenamefont {Wieck}, \citenamefont {Urdampilleta}, \citenamefont {B\"{a}uerle},\ and\ \citenamefont {Meunier}}]{Jadot2021}%
  \BibitemOpen
  \bibfield  {author} {\bibinfo {author} {\bibfnamefont {B.}~\bibnamefont {Jadot}}, \bibinfo {author} {\bibfnamefont {P.-A.}\ \bibnamefont {Mortemousque}}, \bibinfo {author} {\bibfnamefont {E.}~\bibnamefont {Chanrion}}, \bibinfo {author} {\bibfnamefont {V.}~\bibnamefont {Thiney}}, \bibinfo {author} {\bibfnamefont {A.}~\bibnamefont {Ludwig}}, \bibinfo {author} {\bibfnamefont {A.~D.}\ \bibnamefont {Wieck}}, \bibinfo {author} {\bibfnamefont {M.}~\bibnamefont {Urdampilleta}}, \bibinfo {author} {\bibfnamefont {C.}~\bibnamefont {B\"{a}uerle}},\ and\ \bibinfo {author} {\bibfnamefont {T.}~\bibnamefont {Meunier}},\ }\bibfield  {title} {\bibinfo {title} {Distant spin entanglement via fast and coherent electron shuttling},\ }\href {https://doi.org/10.1038/s41565-021-00846-y} {\bibfield  {journal} {\bibinfo  {journal} {Nature Nanotechnology}\ }\textbf {\bibinfo {volume} {16}},\ \bibinfo {pages} {570} (\bibinfo {year} {2021})}\BibitemShut {NoStop}%
\bibitem [{\citenamefont {Bello}\ \emph {et~al.}(2022)\citenamefont {Bello}, \citenamefont {Benito}, \citenamefont {Schuetz}, \citenamefont {Platero},\ and\ \citenamefont {Giedke}}]{Bello2022}%
  \BibitemOpen
  \bibfield  {author} {\bibinfo {author} {\bibfnamefont {M.}~\bibnamefont {Bello}}, \bibinfo {author} {\bibfnamefont {M.}~\bibnamefont {Benito}}, \bibinfo {author} {\bibfnamefont {M.~J.~A.}\ \bibnamefont {Schuetz}}, \bibinfo {author} {\bibfnamefont {G.}~\bibnamefont {Platero}},\ and\ \bibinfo {author} {\bibfnamefont {G.}~\bibnamefont {Giedke}},\ }\bibfield  {title} {\bibinfo {title} {Entangling nuclear spins in distant quantum dots via an electron bus},\ }\href {https://doi.org/10.1103/PhysRevApplied.18.014009} {\bibfield  {journal} {\bibinfo  {journal} {Phys. Rev. Appl.}\ }\textbf {\bibinfo {volume} {18}},\ \bibinfo {pages} {014009} (\bibinfo {year} {2022})}\BibitemShut {NoStop}%
\bibitem [{\citenamefont {Takada}\ \emph {et~al.}(2019)\citenamefont {Takada}, \citenamefont {Edlbauer}, \citenamefont {Lepage}, \citenamefont {Wang}, \citenamefont {Mortemousque}, \citenamefont {Georgiou}, \citenamefont {Barnes}, \citenamefont {Ford}, \citenamefont {Yuan}, \citenamefont {Santos}, \citenamefont {Waintal}, \citenamefont {Ludwig}, \citenamefont {Wieck}, \citenamefont {Urdampilleta}, \citenamefont {Meunier},\ and\ \citenamefont {B\"{a}uerle}}]{Takada2019}%
  \BibitemOpen
  \bibfield  {author} {\bibinfo {author} {\bibfnamefont {S.}~\bibnamefont {Takada}}, \bibinfo {author} {\bibfnamefont {H.}~\bibnamefont {Edlbauer}}, \bibinfo {author} {\bibfnamefont {H.~V.}\ \bibnamefont {Lepage}}, \bibinfo {author} {\bibfnamefont {J.}~\bibnamefont {Wang}}, \bibinfo {author} {\bibfnamefont {P.-A.}\ \bibnamefont {Mortemousque}}, \bibinfo {author} {\bibfnamefont {G.}~\bibnamefont {Georgiou}}, \bibinfo {author} {\bibfnamefont {C.~H.~W.}\ \bibnamefont {Barnes}}, \bibinfo {author} {\bibfnamefont {C.~J.~B.}\ \bibnamefont {Ford}}, \bibinfo {author} {\bibfnamefont {M.}~\bibnamefont {Yuan}}, \bibinfo {author} {\bibfnamefont {P.~V.}\ \bibnamefont {Santos}}, \bibinfo {author} {\bibfnamefont {X.}~\bibnamefont {Waintal}}, \bibinfo {author} {\bibfnamefont {A.}~\bibnamefont {Ludwig}}, \bibinfo {author} {\bibfnamefont {A.~D.}\ \bibnamefont {Wieck}}, \bibinfo {author} {\bibfnamefont {M.}~\bibnamefont {Urdampilleta}}, \bibinfo {author} {\bibfnamefont {T.}~\bibnamefont {Meunier}},\ and\ \bibinfo {author}
  {\bibfnamefont {C.}~\bibnamefont {B\"{a}uerle}},\ }\bibfield  {title} {\bibinfo {title} {Sound-driven single-electron transfer in a circuit of coupled quantum rails},\ }\bibfield  {journal} {\bibinfo  {journal} {Nature Communications}\ }\textbf {\bibinfo {volume} {10}},\ \href {https://doi.org/10.1038/s41467-019-12514-w} {10.1038/s41467-019-12514-w} (\bibinfo {year} {2019})\BibitemShut {NoStop}%
\bibitem [{\citenamefont {Huang}\ and\ \citenamefont {Hu}(2013)}]{Huang2019}%
  \BibitemOpen
  \bibfield  {author} {\bibinfo {author} {\bibfnamefont {P.}~\bibnamefont {Huang}}\ and\ \bibinfo {author} {\bibfnamefont {X.}~\bibnamefont {Hu}},\ }\bibfield  {title} {\bibinfo {title} {Spin qubit relaxation in a moving quantum dot},\ }\href {https://doi.org/10.1103/PhysRevB.88.075301} {\bibfield  {journal} {\bibinfo  {journal} {Phys. Rev. B}\ }\textbf {\bibinfo {volume} {88}},\ \bibinfo {pages} {075301} (\bibinfo {year} {2013})}\BibitemShut {NoStop}%
\bibitem [{\citenamefont {Bertrand}\ \emph {et~al.}(2016)\citenamefont {Bertrand}, \citenamefont {Hermelin}, \citenamefont {Takada}, \citenamefont {Yamamoto}, \citenamefont {Tarucha}, \citenamefont {Ludwig}, \citenamefont {Wieck}, \citenamefont {B\"{a}uerle},\ and\ \citenamefont {Meunier}}]{Bertrand2016}%
  \BibitemOpen
  \bibfield  {author} {\bibinfo {author} {\bibfnamefont {B.}~\bibnamefont {Bertrand}}, \bibinfo {author} {\bibfnamefont {S.}~\bibnamefont {Hermelin}}, \bibinfo {author} {\bibfnamefont {S.}~\bibnamefont {Takada}}, \bibinfo {author} {\bibfnamefont {M.}~\bibnamefont {Yamamoto}}, \bibinfo {author} {\bibfnamefont {S.}~\bibnamefont {Tarucha}}, \bibinfo {author} {\bibfnamefont {A.}~\bibnamefont {Ludwig}}, \bibinfo {author} {\bibfnamefont {A.~D.}\ \bibnamefont {Wieck}}, \bibinfo {author} {\bibfnamefont {C.}~\bibnamefont {B\"{a}uerle}},\ and\ \bibinfo {author} {\bibfnamefont {T.}~\bibnamefont {Meunier}},\ }\bibfield  {title} {\bibinfo {title} {Fast spin information transfer between distant quantum dots using individual electrons},\ }\href {https://doi.org/10.1038/nnano.2016.82} {\bibfield  {journal} {\bibinfo  {journal} {Nature Nanotechnology}\ }\textbf {\bibinfo {volume} {11}},\ \bibinfo {pages} {672} (\bibinfo {year} {2016})}\BibitemShut {NoStop}%
\bibitem [{\citenamefont {Tarucha}\ \emph {et~al.}(1996)\citenamefont {Tarucha}, \citenamefont {Austing}, \citenamefont {Honda}, \citenamefont {van~der Hage},\ and\ \citenamefont {Kouwenhoven}}]{Tarucha1996}%
  \BibitemOpen
  \bibfield  {author} {\bibinfo {author} {\bibfnamefont {S.}~\bibnamefont {Tarucha}}, \bibinfo {author} {\bibfnamefont {D.~G.}\ \bibnamefont {Austing}}, \bibinfo {author} {\bibfnamefont {T.}~\bibnamefont {Honda}}, \bibinfo {author} {\bibfnamefont {R.~J.}\ \bibnamefont {van~der Hage}},\ and\ \bibinfo {author} {\bibfnamefont {L.~P.}\ \bibnamefont {Kouwenhoven}},\ }\bibfield  {title} {\bibinfo {title} {Shell filling and spin effects in a few electron quantum dot},\ }\href {https://doi.org/10.1103/physrevlett.77.3613} {\bibfield  {journal} {\bibinfo  {journal} {Physical Review Letters}\ }\textbf {\bibinfo {volume} {77}},\ \bibinfo {pages} {3613–3616} (\bibinfo {year} {1996})}\BibitemShut {NoStop}%
\bibitem [{\citenamefont {Stotz}\ \emph {et~al.}(2005)\citenamefont {Stotz}, \citenamefont {Hey}, \citenamefont {Santos},\ and\ \citenamefont {Ploog}}]{Stotz2005}%
  \BibitemOpen
  \bibfield  {author} {\bibinfo {author} {\bibfnamefont {J.~A.~H.}\ \bibnamefont {Stotz}}, \bibinfo {author} {\bibfnamefont {R.}~\bibnamefont {Hey}}, \bibinfo {author} {\bibfnamefont {P.~V.}\ \bibnamefont {Santos}},\ and\ \bibinfo {author} {\bibfnamefont {K.~H.}\ \bibnamefont {Ploog}},\ }\bibfield  {title} {\bibinfo {title} {Coherent spin transport through dynamic quantum dots},\ }\href {https://doi.org/10.1038/nmat1430} {\bibfield  {journal} {\bibinfo  {journal} {Nature Materials}\ }\textbf {\bibinfo {volume} {4}},\ \bibinfo {pages} {585–588} (\bibinfo {year} {2005})}\BibitemShut {NoStop}%
\bibitem [{\citenamefont {Suzuki}(1976)}]{Suzuki1976}%
  \BibitemOpen
  \bibfield  {author} {\bibinfo {author} {\bibfnamefont {M.}~\bibnamefont {Suzuki}},\ }\bibfield  {title} {\bibinfo {title} {Generalized trotter{\textquotesingle}s formula and systematic approximants of exponential operators and inner derivations with applications to many-body problems},\ }\href {https://doi.org/10.1007/bf01609348} {\bibfield  {journal} {\bibinfo  {journal} {Communications in Mathematical Physics}\ }\textbf {\bibinfo {volume} {51}},\ \bibinfo {pages} {183} (\bibinfo {year} {1976})}\BibitemShut {NoStop}%
\bibitem [{\citenamefont {Hatano}\ and\ \citenamefont {Suzuki}(2005)}]{Hatano2005}%
  \BibitemOpen
  \bibfield  {author} {\bibinfo {author} {\bibfnamefont {N.}~\bibnamefont {Hatano}}\ and\ \bibinfo {author} {\bibfnamefont {M.}~\bibnamefont {Suzuki}},\ }\bibfield  {title} {\bibinfo {title} {Finding exponential product formulas of higher orders},\ }in\ \href {https://doi.org/10.1007/11526216_2} {\emph {\bibinfo {booktitle} {Quantum Annealing and Other Optimization Methods}}}\ (\bibinfo  {publisher} {Springer Berlin Heidelberg},\ \bibinfo {year} {2005})\ pp.\ \bibinfo {pages} {37--68}\BibitemShut {NoStop}%
\bibitem [{\citenamefont {{Bychkov}}\ and\ \citenamefont {{Rashba}}(1984)}]{Rashba1984}%
  \BibitemOpen
  \bibfield  {author} {\bibinfo {author} {\bibfnamefont {Y.~A.}\ \bibnamefont {{Bychkov}}}\ and\ \bibinfo {author} {\bibfnamefont {{\'E}.~I.}\ \bibnamefont {{Rashba}}},\ }\bibfield  {title} {\bibinfo {title} {{Properties of a 2D electron gas with lifted spectral degeneracy}},\ }\href@noop {} {\bibfield  {journal} {\bibinfo  {journal} {Soviet Journal of Experimental and Theoretical Physics Letters}\ }\textbf {\bibinfo {volume} {39}},\ \bibinfo {pages} {78} (\bibinfo {year} {1984})}\BibitemShut {NoStop}%
\bibitem [{\citenamefont {Dresselhaus}(1955)}]{Dresselhaus1955}%
  \BibitemOpen
  \bibfield  {author} {\bibinfo {author} {\bibfnamefont {G.}~\bibnamefont {Dresselhaus}},\ }\bibfield  {title} {\bibinfo {title} {Spin-orbit coupling effects in zinc blende structures},\ }\href {https://doi.org/10.1103/PhysRev.100.580} {\bibfield  {journal} {\bibinfo  {journal} {Phys. Rev.}\ }\textbf {\bibinfo {volume} {100}},\ \bibinfo {pages} {580} (\bibinfo {year} {1955})}\BibitemShut {NoStop}%
\bibitem [{\citenamefont {Studer}\ \emph {et~al.}(2009)\citenamefont {Studer}, \citenamefont {Salis}, \citenamefont {Ensslin}, \citenamefont {Driscoll},\ and\ \citenamefont {Gossard}}]{Studer2009}%
  \BibitemOpen
  \bibfield  {author} {\bibinfo {author} {\bibfnamefont {M.}~\bibnamefont {Studer}}, \bibinfo {author} {\bibfnamefont {G.}~\bibnamefont {Salis}}, \bibinfo {author} {\bibfnamefont {K.}~\bibnamefont {Ensslin}}, \bibinfo {author} {\bibfnamefont {D.~C.}\ \bibnamefont {Driscoll}},\ and\ \bibinfo {author} {\bibfnamefont {A.~C.}\ \bibnamefont {Gossard}},\ }\bibfield  {title} {\bibinfo {title} {Gate-controlled spin-orbit interaction in a parabolic gaas/algaas quantum well},\ }\bibfield  {journal} {\bibinfo  {journal} {Physical Review Letters}\ }\textbf {\bibinfo {volume} {103}},\ \href {https://doi.org/10.1103/physrevlett.103.027201} {10.1103/physrevlett.103.027201} (\bibinfo {year} {2009})\BibitemShut {NoStop}%
\bibitem [{\citenamefont {Dettwiler}\ \emph {et~al.}(2017)\citenamefont {Dettwiler}, \citenamefont {Fu}, \citenamefont {Mack}, \citenamefont {Weigele}, \citenamefont {Egues}, \citenamefont {Awschalom},\ and\ \citenamefont {Zumb\"uhl}}]{Dettwiler2017}%
  \BibitemOpen
  \bibfield  {author} {\bibinfo {author} {\bibfnamefont {F.}~\bibnamefont {Dettwiler}}, \bibinfo {author} {\bibfnamefont {J.}~\bibnamefont {Fu}}, \bibinfo {author} {\bibfnamefont {S.}~\bibnamefont {Mack}}, \bibinfo {author} {\bibfnamefont {P.~J.}\ \bibnamefont {Weigele}}, \bibinfo {author} {\bibfnamefont {J.~C.}\ \bibnamefont {Egues}}, \bibinfo {author} {\bibfnamefont {D.~D.}\ \bibnamefont {Awschalom}},\ and\ \bibinfo {author} {\bibfnamefont {D.~M.}\ \bibnamefont {Zumb\"uhl}},\ }\bibfield  {title} {\bibinfo {title} {Stretchable persistent spin helices in gaas quantum wells},\ }\href {https://doi.org/10.1103/PhysRevX.7.031010} {\bibfield  {journal} {\bibinfo  {journal} {Phys. Rev. X}\ }\textbf {\bibinfo {volume} {7}},\ \bibinfo {pages} {031010} (\bibinfo {year} {2017})}\BibitemShut {NoStop}%
\bibitem [{\citenamefont {Kn\"{o}rzer}\ \emph {et~al.}(2018)\citenamefont {Kn\"{o}rzer}, \citenamefont {Schuetz}, \citenamefont {Giedke}, \citenamefont {Huebl}, \citenamefont {Weiler}, \citenamefont {Lukin},\ and\ \citenamefont {Cirac}}]{Knrzer2018}%
  \BibitemOpen
  \bibfield  {author} {\bibinfo {author} {\bibfnamefont {J.}~\bibnamefont {Kn\"{o}rzer}}, \bibinfo {author} {\bibfnamefont {M.~J.~A.}\ \bibnamefont {Schuetz}}, \bibinfo {author} {\bibfnamefont {G.}~\bibnamefont {Giedke}}, \bibinfo {author} {\bibfnamefont {H.}~\bibnamefont {Huebl}}, \bibinfo {author} {\bibfnamefont {M.}~\bibnamefont {Weiler}}, \bibinfo {author} {\bibfnamefont {M.~D.}\ \bibnamefont {Lukin}},\ and\ \bibinfo {author} {\bibfnamefont {J.~I.}\ \bibnamefont {Cirac}},\ }\bibfield  {title} {\bibinfo {title} {Solid-state magnetic traps and lattices},\ }\bibfield  {journal} {\bibinfo  {journal} {Physical Review B}\ }\textbf {\bibinfo {volume} {97}},\ \href {https://doi.org/10.1103/physrevb.97.235451} {10.1103/physrevb.97.235451} (\bibinfo {year} {2018})\BibitemShut {NoStop}%
\end{thebibliography}%

\end{document}